\documentclass[11pt,a4paper]{article}
\pdfoutput=1

\usepackage{jheppub}
\usepackage{amsmath}
\usepackage{amssymb}
\usepackage{graphics}
\usepackage[active]{srcltx}
\usepackage{pdfsync}
\usepackage{shuffle}
\usepackage{slashed}
\usepackage{hyperref}
\usepackage{subfigure}

\newcommand{\bea}{\begin{eqnarray}}
\newcommand{\eea}{\end{eqnarray}}
\newcommand{\nn}{\nonumber \\}

\def\W #1{\widetilde{#1}}

\def\eref#1{(\ref{#1})}

\def\a{{\alpha}}

\def\b{{\beta}}

\def\la{\lambda}

\allowdisplaybreaks

\title{Note on hidden zeros and expansions of tree-level amplitudes}
\author[a]{Hao Huang} \author[a]{Ye Yang} \author[a]{Kang Zhou}

\affiliation[a]{Center for Gravitation and Cosmology, College of Physical Science and Technology, Yangzhou University,\\
No.180, Siwangting Road, Yangzhou, 225009, P.R. China}

\emailAdd{yzudrugster@163.com} \emailAdd{18550005886@163.com} \emailAdd{zhoukang@yzu.edu.cn}

\date{\today}
\abstract{In this note, we derive and interpret hidden zeros of tree-level amplitudes of various theories, including Yang-Mills, non-linear sigma model, special Galileon, Dirac-Born-Infeld, and gravity, by utilizing universal expansions of tree-level amplitudes which expand amplitudes of different theories to those of bi-adjoint scalar theory. Hidden zeros of above amplitudes are attributed to zeros of bi-adjoint scalar amplitudes which can be easily proved. For unordered amplitudes such as gravitational ones,
the kinematic condition for zeros causes potential divergences of propagators. We also show the detailed mechanism of eliminating such divergences.}


\begin{document}

\maketitle \flushbottom

\section{Introduction}
\label{sec-intro}

The modern researches on S-matrix have exposed marvelous properties of scattering amplitudes which are not evident upon inspecting traditional
Feynman rules. The last few decades have witnessed not only the discovery of significant new structures such as color-kinematic duality and double copy construction \cite{Bern:2008qj,Bern:2010ue,Bern:2019prr}, but also the establishing of novel formalisms of amplitudes, for instance the well known Cachazo-He-Yuan (CHY) formula \cite{Cachazo:2013hca,Cachazo:2013iea,Cachazo:2014nsa,Cachazo:2014xea}, and the geometric and combinatorial description \cite{Arkani-Hamed:2017mur,Arkani-Hamed:2023lbd,Arkani-Hamed:2023mvg,Arkani-Hamed:2023jry,Arkani-Hamed:2024nhp}.
Very recently, an amazing new property of tree-level amplitudes called hidden zeros, as well as novel factorizations near such zeros, are discovered in \cite{Arkani-Hamed:2023swr}. By employing the kinematic mesh and stringy curve integrals, they found the tree amplitudes for certain theories, including the ${\rm Tr}(\phi^3)$ theory, non-linear sigma model (NLSM) and Yang-Mills (YM) theory, vanish on special loci in the kinematic space. By turning on one of vanished Mandelstam variables to be non-zero, each amplitude factorizes into three amputated currents. Subsequently, such interesting hidden zeros have been further studied in \cite{Rodina:2024yfc,Bartsch:2024amu,Li:2024qfp,Zhang:2024iun,Zhou:2024ddy,Zhang:2024efe}, especially have been extended to unordered tree amplitudes such as special Galileon (SG), Dirac-Born-Infeld (DBI), and gravitational (GR) ones. The related new factorizations, without taking residue at any pole, also caused enormous investigations, see in \cite{Cao:2024gln,Arkani-Hamed:2024fyd,Cao:2024qpp,Zhang:2024iun,Zhou:2024ddy,Zhang:2024efe}.

Since the hidden zeros hold for a wide range of theories, with the similar constraints on kinematics, it is natural to expect that such an universal behavior can be interpreted via an unique framework. Unfortunately, the original method in \cite{Arkani-Hamed:2023swr}, can not be generalized to unordered amplitudes by using our current knowledge of underlying geometric and combinatorial structure. The first unified understanding for hidden zeros among different theories was proposed in \cite{Zhang:2024efe}, by exploiting the powerful CHY formula. In \cite{Zhang:2024efe}, hidden zeros for tree-level amplitudes of ${\rm Tr}(\phi^3)$, NLSM, YM, SG, DBI and GR theories are attributed to the divergence of Jacobian appears in the measure of CHY contour integral.

In this short note, we propose an alternative unified interpretation for hidden zeros, base on the universal expansions of amplitudes which expand amplitudes of a variety of theories to bi-adjoint scalar (BAS) ones, with polynomial coefficients depend on kinematics carried by external particles \cite{Stieberger:2016lng,Schlotterer:2016cxa,Chiodaroli:2017ngp,Nandan:2016pya,delaCruz:2016gnm,
Fu:2017uzt,Teng:2017tbo,Du:2017kpo,Du:2017gnh,Feng:2019tvb,Zhou:2019gtk,Zhou:2019mbe,Wei:2023yfy,Hu:2023lso,Du:2024dwm,Zhou:2024qwm,Zhou:2024qjh}.
Comparing with the method in \cite{Zhang:2024efe}, our approach leads to a totally different picture: by applying the well known Kleiss-Kuijf (KK) relation, hidden zeros are ultimately attributed to the vanishing of a certain set of BAS amplitudes, for special loci in the kinematic space. Thus, these zeros of BAS amplitudes play the basic role in our scheme. As will be seen, proving such zeros of BAS amplitudes is much more easier than proving zeros of other theories.

An important motivation for this note, is to understand the detailed mechanism of eliminating potential divergences in unordered amplitudes. Let us use hidden zeros of GR amplitudes to demonstrate such divergences. For an $n$-point GR amplitude, one can pick up two external particles $i$ and $j$, and split the remaining $n-2$ into two non-empty sets $A$ and $B$. Then we impose the condition
\bea
s_{ab}=0\,,~~~~{\rm with}~a\in A\,,~b\in B\,,~~\label{kinematic-condi-0-phi3}
\eea
as well as the similar constraint on polarizations. Here and after, the Mandelstam variable $s_{p\cdots q}$ is defined as usual
\bea
s_{p\cdots q}\equiv k_{p\cdots q}^2\,,~~~{\rm with}~k_{p\cdots q}\equiv\sum_{\ell=p}^q\,k_\ell\,,~~~~\label{mandelstam}
\eea
where $k_\ell$ is the momentum carried by the external leg $\ell$. As proved in \cite{Zhang:2024efe} and will be reproduced in this note, the GR amplitude vanish under the above constraints. Base on the kinematic condition \eref{kinematic-condi-0-phi3}, a straightforward observation is, the divergent propagators in the form $1/s_{ab}$ can enter the GR amplitude, since two external legs $a$ and $b$ of an unordered amplitude are allowed to be nearby in corresponding Feynman diagrams. Thus, the statement of zeros makes sense if and only if all these divergences are canceled/removed. We emphasize that the proof of zeros based on CHY formula is strict, and will not be spoiled by above divergences. However, that proof can not tell us how the divergences arise from $1/s_{ab}$ are eliminated. The universal expansions offer an useful tool for considering the mechanism of removing divergences, since propagators $1/s_{ab}$ are explicit contained in the BAS basis in expansions, and the cancellation of them are governed by relations among BAS amplitudes with different orderings, as well as the corresponding polynomial coefficients. Therefore, in this note we will investigate the elimination of such divergences in detail.

The remainder of this note is organized as follows. In section \ref{sec-background}, we rapidly review the necessary background including hidden zeros of tree-level BAS amplitudes, as well as universal expansions of tree-level amplitudes for different theories. Then, in section \ref{sec-YM-NLSM}, we reproduce hidden zeros of ordered tree-level amplitudes, including YM and NLSM ones, by utilizing universal expansions and zeros of BAS amplitudes. In section \ref{sec-unordered}, we continue to show hidden zeros of unordered amplitudes including SG, DBI and GR ones, by applying the similar method, and consider the removing of divergences case by case. We end with a brief summary and discussion in section \ref{sec-summary}. In appendix \ref{sec-appen}, we give a simple proof for hidden zeros of BAS amplitudes, by exploiting solely traditional Feynman rules. In appendix \ref{sec-appen2}, some cancellations of divergent terms in the GR case are shown.

\section{Background}
\label{sec-background}

For readers' convenience, in this section we give a brief review for the background. In subsection \ref{subsec-BAS}, we rapidly introduce the tree-level BAS amplitudes, as well as their hidden zeros. In subsection \ref{subsec-expansion}, we review the universal expansions of tree-level amplitudes of various theories, including YM, NLSM, SG, BI and GR.

\subsection{Tree-level BAS amplitudes and their hidden zeros}
\label{subsec-BAS}

The bi-adjoint scalar (BAS) theory describes the cubic interaction of massless scalars, with the Lagrangian
\bea
{\cal L}^{\rm BAS}={1\over2}\,\partial_\mu\phi^{Aa}\,\partial^{\mu}\phi^{Aa}+{\lambda\over3!}\,F^{ABC}f^{abc}\,
\phi^{Aa}\phi^{Bb}\phi^{Cc}\,,~~\label{Lag-BAS}
\eea
where $F^{ABC}={\rm tr}([T^A,T^B]T^C)$ and $f^{abc}={\rm tr}([T^a,T^b]T^c)$ are structure constants of two Lie groups respectively, and each scalar field $\phi^{Aa}$ carries two group indices.
The usual technic of decomposing group factors turns a tree-level BAS amplitude to
\bea
A^{\rm BAS}_n=\sum_{\sigma\in{\cal S}_n\setminus Z_n}\,\sum_{\sigma'\in{\cal S}'_n\setminus Z'_n}\,
{\rm tr}[T^{A_{\sigma_1}},\cdots T^{A_{\sigma_n}}]\,{\rm tr}[T^{a_{\sigma'_1}}\cdots T^{a_{\sigma'_n}}]\,
{\cal A}^{\rm BAS}_n(\sigma_1,\cdots,\sigma_n|\sigma'_1,\cdots,\sigma'_n)\,,
\eea
where $A_n$ represents the full $n$-point tree-level amplitude with coupling constants stripped off. The summation is for all cyclically inequivalent orderings among external scalars, encoded by un-cyclic permutations ${\cal S}_n\setminus Z_n$ and ${\cal S}'_n\setminus Z'_n$ respectively. Each partial amplitude ${\cal A}^{\rm BAS}_n(\sigma_1, \cdots, \sigma_n|\sigma'_1, \cdots, \sigma'_n)$ consist solely of propagators for massless scalars, and is simultaneously planar with respect to two orderings. For instance, the $4$-point amplitude ${\cal A}^{\rm BAS}_4(1,2,3,4|1,2,4,3)$ reads
\bea
{\cal A}^{\rm BAS}_4(1,2,3,4|1,2,4,3)={1\over s_{12}}\,,
\eea
up to an overall $\pm$ sign. In this example, the amplitude only involves the propagator $1/s_{12}$,
since the Feynman diagrams correspond to $1/s_{14}$ and $1/s_{13}$ are not compatible with two orderings $(1,2,3,4)$ and $(1,2,4,3)$ simultaneously.
The anti-symmetry of structure constants $F^{ABC}$ and $f^{abc}$ indicates an overall sign $\pm$ carried by each partial amplitude,
which arises from swapping lines on vertices. This sign can be determined by counting flips of external legs, as detailed in \cite{Cachazo:2013iea}.
For convenience, sometimes we denote an $n$-point partial amplitude as ${\cal A}^{\rm BAS}_n({\pmb\sigma}_n|{\pmb\sigma}'_n)$, where ${\pmb\sigma}_n$ and ${\pmb\sigma}'_n$ stand for two orderings.

A certain set of partial BAS amplitudes vanish on special loci \eref{kinematic-condi-0-phi3} in kinematic space, such behavior is called the hidden zeros of BAS amplitudes. More explicitly, if two orderings ${\pmb\sigma}_n$ and ${\pmb\sigma}'_n$ carried by ${\cal A}_{\rm BAS}({\pmb\sigma}_n|{\pmb\sigma}'_n)$ satisfy
\bea
{\pmb\sigma}_n=\pmb A,i,\pmb B,j\,,~~~~{\pmb\sigma}'_n=\pmb A',i,\pmb B',j\,,~~~~{\rm up~to~cyclic~permutations}\,,~~\label{compa-order}
\eea
where $\pmb A$ and $\pmb A'$ are two ordered sets obtained by introducing orderings to elements in $A$, while $\pmb B$ and $\pmb B'$ are ordered sets obtained by giving orderings to elements in $B$, then the amplitude behaves as
\bea
{\cal A}^{\rm BAS}_n(\pmb A,i,\pmb B,j|\pmb A',i,\pmb B',j)&\xrightarrow[]{\eref{kinematic-condi-0-phi3}}&0\,.~~\label{zero-BAS}
\eea
As in the above, in this paper we use bold letters to denote ordered sets.
For latter convenience, we call two orderings in \eref{compa-order} as orderings which are compatible with the kinematic condition \eref{kinematic-condi-0-phi3}. In other words, a compatible ordering is that elements in $A$ and $B$ are separated by legs $i$ and $j$. Hidden zeros in \eref{zero-BAS} can be derived via a variety of methods, such as the kinematic mesh, stringy curve integrals, CHY formula, and so on \cite{Arkani-Hamed:2023swr,Bartsch:2024amu,Li:2024qfp,Zhang:2024iun}. The above novel formulas of scattering amplitudes may be unfamiliar for various readers, thus, in appendix \ref{sec-appen} we give the proof of \eref{zero-BAS} proposed in \cite{Zhou:2024ddy}, by utilizing traditional Feynman rules.

\subsection{Expansions of tree-level amplitudes}
\label{subsec-expansion}

As studied in \cite{Stieberger:2016lng,Schlotterer:2016cxa,Chiodaroli:2017ngp,Nandan:2016pya,delaCruz:2016gnm,
Fu:2017uzt,Teng:2017tbo,Du:2017kpo,Du:2017gnh,Feng:2019tvb,Zhou:2019gtk,Zhou:2019mbe,Wei:2023yfy,Hu:2023lso,Du:2024dwm,Zhou:2024qwm,Zhou:2024qjh},
tree-level amplitudes for a wide range of theories can be expanded to BAS ones. According to the work in this note, in this subsection we only introduce such
expansions for YM, NLSM, GR, SG and BI amplitudes.

The expansions of YM and GR amplitudes are given as
\bea
{\cal A}^{\rm YM}_n(\pmb\sigma_n)=\sum_{\pmb\a_{n-2}}\,C^\epsilon(\pmb\a_{n-2})\,{\cal A}^{\rm BAS}_n(i,\pmb\a_{n-2},j|\pmb\sigma_n)\,,
~~\label{expan-YM}
\eea
and
\bea
{\cal A}^{\rm GR}_n=\sum_{\pmb\a_{n-2}}\,\sum_{\pmb\b_{n-2}}\,C^\epsilon(\pmb\a_{n-2})\,C^{\W\epsilon}(\pmb\b_{n-2})\,
{\cal A}^{\rm BAS}_n(i,\pmb\a_{n-2},j|i,\pmb\b_{n-2},j)\,,~~\label{expan-GR}
\eea
where $\pmb\a_{n-2}$ and $\pmb\b_{n-2}$ are ordered sets for $n-2$ elements in $\{1,\cdots,n\}\setminus\{i,j\}$.
Legs $i$ and $j$ fixed at two ends of the ordering can be chosen arbitrary. Here we use the letters $i$ and $j$ to denote them,
since we will chose them as $i$ and $j$ in \eref{zero-BAS} when considering zeros of YM and GR amplitudes.
The polarization tensor of an external graviton is decomposed as $\varepsilon_\ell^{\mu\nu}=\epsilon_\ell^\mu\W\epsilon_\ell^\nu$, where $\epsilon_\ell^\mu$ and $\W\epsilon_\ell^\mu$ are two polarization vectors. For the pure Einstein gravity, $\epsilon_\ell^\mu$ is the same as $\W\epsilon_\ell^\mu$. For the extended model that Einstein gravity coupled to $B$-field and dilaton, they are independent of each other.
In \eref{expan-GR}, each $C^\epsilon(\pmb\a_{n-2})$ is independent of $\W\epsilon_\ell$, while each $C^{\W\epsilon}(\pmb\b_{n-2})$ is independent of
$\epsilon_\ell$, with $\ell\in\{1,\cdots,n\}$. Combining \eref{expan-GR} and \eref{expan-YM} together leads to the expansion of GR amplitudes to YM ones,
\bea
{\cal A}^{\rm GR}_n=\sum_{\pmb\b_{n-2}}\,C^{\W\epsilon}(\pmb\b_{n-2})\,
{\cal A}^{\rm YM}_n(i,\pmb\b_{n-2},j)\,.~~\label{expan-GRtoYM}
\eea

The coefficients $C^\epsilon(\pmb\a_{n-2})$ (or $C^{\W\epsilon}(\pmb\b_{n-2})$) can be constructed as follows. For a given ordering $(i,\pmb\a_{n-2},j)$, with $\pmb\a_{n-2}=\{\a_1,\cdots,\a_{n-2}\}$,
let us represent the ordering
as $i\dot{<} \a_1\dot{<}\cdots\dot{<} \a_{n-2}\dot{<}j$. We also chose a reference ordering $g_1\prec g_2\prec\cdots\prec g_{n-2}$,
where each $g_\ell$ with $\ell\in\{1,\cdots,n-2\}$ is an element in $\pmb\a_{n-2}$. This reference ordering is labeled by $\pmb{\cal R}$.
The central procedure of the algorithm is to construct all appropriate ordered splittings compatible with
the given ordering $i\dot{<} \a_1\dot{<}\cdots\dot{<} \a_{n-2}\dot{<}j$, via the following procedure,
\begin{itemize}
\item The first step is to construct all possible ordered sets $\pmb r_0=\{r^0_1,\cdots,r^0_{|0|}\}$ satisfying two conditions, (1) each element in $\pmb r_0$ belong to $\pmb\a_{n-2}$, (2) $r^0_1\dot{<}r^0_2\dot{<}\cdots\dot{<}r^0_{|0|}$, respecting to the ordering $i\dot{<} \a_1\dot{<}\cdots\dot{<} \a_{n-2}\dot{<}j$.
    Here and after, $|\ell|$ stands for the length of the set $\pmb r_\ell$.
    Notice that $\pmb r_0$ is allowed to be empty.
\item For each $\pmb r_0$, we eliminate its elements in $(i,\pmb\a_{n-2},j)$ and $\pmb{\cal R}$, resulting in two new ordered sets $\pmb\a_{n-2}\setminus r_0$ and $\pmb{\cal R}\setminus r_0$, where $r_0$ denotes the unordered set generated from $\pmb r_0$ by stripping off the ordering. Suppose $R_1$ is the lowest element in the reduced reference ordering $\pmb{\cal R}\setminus r_0$, we construct all possible ordered sets $\pmb r_1$ as $\pmb r_1=\{r_1^1,\cdots,r_{|1|-1}^1,R_1\}$, satisfying
    (1) each element in $\pmb r_1$ belong to $\pmb\a_0\setminus r_0$, (2) $r_1^1\dot{<}r_2^1\dot{<}\cdots\dot{<}r_{|1|-1}^1\dot{<}R_1$, respecting to the ordering $\pmb\a_{n-2}\setminus r_0$.
\item Continue to construct $\pmb r_2=\{r^2_1,\cdots,r^2_{|2|-1},R_2\}$, $\pmb r_3=\{r^3_1,\cdots,r^3_{|3|-1},R_3\}$, and so on, by iterating the second step, until $ r_0\cup r_1\cup\cdots\cup r_f=\{1,\cdots,n\}$.
\end{itemize}
Each ordered splitting is understood as the corresponding ordered set $\{\pmb r_0,\pmb r_1,\cdots,\pmb r_f\}$, where each ordered set
$\pmb r_k$ serves as an element.

For a given ordered splitting, the ordered set $\pmb r_0$ corresponds to the kinematic factor
\bea
F_{\pmb r_0}=\epsilon_j\cdot f_{r^0_{|0|}}\cdot f_{r^0_{|0|-1}}\cdots f_{r^0_1}\cdot\epsilon_i\,,~~\label{F0}
\eea
where $f_\ell^{\mu\nu}\equiv k_\ell^\mu\epsilon_\ell^\nu-\epsilon_\ell^\mu k_\ell^\nu$, and $\epsilon_\ell^\mu$ is the polarization vector carried by the external particle $\ell$.
For an ordered set $\pmb r_k$ with $k\neq 0$, the corresponding kinematic factor is given as
\bea
F_{\pmb r_k}=\epsilon_{R_\ell}\cdot f_{r^k_{|k|-1}}\cdots f_{r^k_1}\cdot Z_{r^k_1}\,.~~\label{Fl}
\eea
In the above, the combinatorial momentum is defined as $Z_{r^k_1}=k_i+\W Z_{r^k_1}$, where $\W Z_{r^k_1}$ is the summation of momenta of external legs satisfying two conditions: (1) legs
at the l.h.s of $r^k_1$ in the ordering $(i,\pmb\a_{n-2},j)$, (2) legs belong to $\pmb r_h$ at the l.h.s of $\pmb r_k$
in the ordered splitting, namely, $h<k$. Then, the coefficient $C^\epsilon(\pmb\a_{n-2})$ is constructed as
\bea
C^\epsilon(\pmb\a_{n-2})=\sum_{\rm splitting}\,\prod_{k=0}^f\,F_{\pmb r_k}\,,~~\label{c-YM}
\eea
where the summation is among ordered splittings with respect to the ordering $(i,\pmb\a_{n-2},j)$.

To be more concrete, let us explicitly give the construction for $C^\epsilon(3,2)$, with the ordering $1\dot{<}3\dot{<}2\dot{<}4$, where $i=1$ and $j=4$. The reference ordering is chosen to be $4\prec3\prec2\prec1$. The candidates of $\pmb r_0$, respecting the ordering $1\dot{<}3\dot{<}2\dot{<}4$, are found as $\emptyset$, $\{2\}$, $\{3\}$, $\{3,2\}$. For $\pmb r_0=\emptyset$, the lowest element in the reduced reference ordering $3\prec2$ is $3$, then $\pmb r_1$ which is compatible with the ordering $1\dot{<}3\dot{<}2\dot{<}4$ is forced to be $\pmb r_1=\{3\}$. Consequently, we obtain the ordered splitting $\{\{1,4\},\{3\},\{2\}\}$ for $\pmb r_0=\emptyset$. Similarly, one can find $\{\{1,2,4\},\{3\}\}$, $\{\{1,3,4\},\{2\}\}$ and $\{\{1,3,2,4\}\}$ for remaining three choices of $\pmb r_0$. The coefficient $C^\epsilon(2,3)$ is then determined by substituting kinematic factors in \eref{F0} and \eref{Fl},
\bea
C^\epsilon(3,2)=(\epsilon_4\cdot\epsilon_1)(\epsilon_3\cdot k_1)(\epsilon_2\cdot k_{13})+(\epsilon_4\cdot f_2\cdot\epsilon_1)(\epsilon_3\cdot k_1)+(\epsilon_4\cdot f_3\cdot\epsilon_1)(\epsilon_2\cdot k_{13})+(\epsilon_4\cdot f_2\cdot f_3\cdot\epsilon_1)\,.
\eea

Similarly, the $U(N)$ NLSM and SG amplitudes can be expanded as
\bea
{\cal A}^{\rm NLSM}_n(\pmb\sigma_n)=\sum_{\pmb\a_{n-2}}\,\hat{C}(\pmb\a_{n-2})\,{\cal A}^{\rm BAS}_n(i,\pmb\a_{n-2},j|\pmb\sigma_n)\,,
~~\label{expan-NLSM}
\eea
and
\bea
{\cal A}^{\rm SG}_n=\sum_{\pmb\a_{n-2}}\,\sum_{\pmb\b_{n-2}}\,\hat{C}(\pmb\a_{n-2})\,\hat C(\pmb\b_{n-2})\,
{\cal A}^{\rm BAS}_n(i,\pmb\a_{n-2},j|i,\pmb\b_{n-2},j)\,.~~\label{expan-SG}
\eea
Combining \eref{expan-NLSM} and \eref{expan-SG} together gives the expansion of SG amplitudes into $U(N)$ NLSM ones,
\bea
{\cal A}^{\rm SG}_n=\sum_{\pmb\b_{n-2}}\,\hat C(\pmb\b_{n-2})\,
{\cal A}^{\rm NLSM}_n(i,\pmb\b_{n-2},j)\,.~~\label{expan-SGtoNLSM}
\eea
It is worth to clarify the explicit definitions of $U(N)$ NLSM and SG theories. The $U(N)$ NLSM theory under consideration has the following Lagrangian in Cayley parametrization,
\bea
{\cal L}^{{\tiny\mbox{NLSM}}}={1\over 8\la^2}{\rm Tr}(\partial_\mu {\rm U}^\dag\partial^\mu {\rm U})\,,
\eea
where
\bea
{\rm U}=(\mathbb{I}+\lambda \Phi)\,(\mathbb{I}-\lambda\Phi)^{-1}\,,~~~\Phi=\phi_I T^I\,.~~\label{U}
\eea
Here $\mathbb{I}$ stands for the identity matrix, while $T^I$'s are generators of $U(N)$.
The general pure Galileon Lagrangian is \cite{Nicolis:2008in}
\bea
{\cal L}^{\tiny\mbox{SG}}=-{1\over2}\,\partial_\mu\phi\,\partial^\mu\phi+\sum_{m=3}^\infty g_m\,{\cal L}_m\,,
\eea
with
\bea
{\cal L}_m=\phi\,{\rm det}\,\{\partial^{\mu_i}\,\partial_{\nu_j}\phi\}_{i,j=1}^{m-1}\,.
\eea
The SG theory is that all tree-level amplitudes with odd numbers
of external particles vanish, due to constrains on coupling constants $g_m$ \cite{Cachazo:2014xea}.

In the \eref{expan-NLSM} and \eref{expan-SG}, each coefficient $\hat C(\pmb\a_{n-2})$ (or $\hat C(\pmb\b_{n-2})$) is given as
\bea
\hat C(\pmb\a_{n-2})=\prod_{\ell=1}^{n-2}\,k_{\a_\ell}\cdot X_{\a_\ell}\,,~~\label{c-NLSM}
\eea
for the given ordered set $\pmb\a_{n-2}=\{\a_1,\cdots,\a_{n-2}\}$. Here the combinatorial momentum $X_{\a_\ell}$ is the summation for momentum of external legs at the l.h.s of $\a_\ell$ in the ordering $(i,\pmb\a_{n-2},j)$. For instance, suppose the ordering is $(1,2,3,4)$, then $X_2=k_1$, $X_3=k_{12}$.

The combination of coefficients in \eref{c-YM} and \eref{c-NLSM} lead to the expansion of Born-Infeld (BI) amplitudes to BAS ones
\bea
{\cal A}^{\rm BI}_n=\sum_{\pmb\a_{n-2}}\,\sum_{\pmb\b_{n-2}}\,C^\epsilon(\pmb\a_{n-2})\,\hat C(\pmb\b_{n-2})\,
{\cal A}^{\rm BAS}_n(i,\pmb\a_{n-2},j|i,\pmb\b_{n-2},j)\,.~~\label{expan-BI}
\eea
The corresponding Lagrangian of this theory is given by \cite{Tseytlin:1999dj}
\bea
{\cal L}^{\tiny\mbox{BI}}=\ell^{-2}\,\Big(\sqrt{-{\rm det}(\eta_{\mu\nu}-\ell\, F_{\mu\nu})}-1\Big)\,.~~\label{Lag-BI}
\eea
Summing over $\pmb\a_{n-2}$ or $\pmb\b_{n-2}$ in \eref{expan-BI}, we get expansions of BI amplitudes to YM or NLSM ones,
\bea
{\cal A}^{\rm BI}_n=\sum_{\pmb\b_{n-2}}\,\hat C(\pmb\b_{n-2})\,
{\cal A}^{\rm YM}_n(i,\pmb\b_{n-2},j)\,,~~\label{expan-BItoYM}
\eea
and
\bea
{\cal A}^{\rm BI}_n=\sum_{\pmb\a_{n-2}}\,C^\epsilon(\pmb\a_{n-2})\,
{\cal A}^{\rm NLSM}_n(i,\pmb\a_{n-2},j)\,.~~\label{expan-BItoNLSM}
\eea
%

\section{Hidden zeros of ordered amplitudes}
\label{sec-YM-NLSM}

In this section, we show hidden zeros of tree-level YM and NLSM amplitudes, by exploiting universal expansions introduced in section \ref{subsec-expansion}, as well as hidden zeros of certain BAS amplitudes in \eref{zero-BAS}. In the original expansions of YM and NLSM amplitudes in \eref{expan-YM} and \eref{expan-NLSM}, the orderings carried by BAS amplitudes are not compatible with the kinematic condition \eref{kinematic-condi-0-phi3}. Thus, generally speaking, BAS amplitudes in \eref{expan-YM} and \eref{expan-NLSM} do not vanish as in \eref{zero-BAS}. However, as we will shown, the special loci in kinematic space force BAS amplitudes to be separated into different sets, while amplitudes in each set share exactly the same coefficient. Then one can apply the Kleiss-Kuijf (KK) relation to convert BAS amplitudes in each given set to those carry orderings compatible with the kinematic condition \eref{kinematic-condi-0-phi3}, and use zeros in \eref{zero-BAS} to argue the vanishing of target amplitudes. The procedure described above is the main idea in subsequent subsections.

\subsection{YM amplitudes}
\label{subsec-YM}

The tree-level YM amplitudes ${\cal A}^{\rm YM}_n(\pmb A,i,\pmb B,j)$, which carry the orderings compatible with the kinematic condition \eref{kinematic-condi-0-phi3}, namely legs in $A$ and $B$ are separated by legs $i$ and $j$, exhibit the following hidden zeros \cite{Arkani-Hamed:2023swr},
\bea
{\cal A}^{\rm YM}_n(\pmb A,i,\pmb B,j)&\xrightarrow[]{\eref{kinematic-condi-0-phi3},\eref{kinematic-condi-0-YM}}&0\,,~~\label{zero-YM}
\eea
where the polarization vectors are constrained by
\bea
\epsilon_a\cdot\epsilon_b=0\,,~~~~\epsilon_a\cdot k_b=0\,,~~~~\epsilon_b\cdot k_a=0\,,~~~~{\rm with}~a\in A\,,~b\in B\,.~~\label{kinematic-condi-0-YM}
\eea
This subsection aims to show the behavior in \eref{zero-YM}, by utilizing the expansion in \eref{expan-YM}.

We choose legs $i$ and $j$ in \eref{expan-YM} to be the same as $i$ and $j$ in \eref{zero-YM}, then the YM amplitude ${\cal A}^{\rm YM}_n(\pmb A,i,\pmb B,j)$ under consideration is expanded as
\bea
{\cal A}^{\rm YM}_n(\pmb A,i,\pmb B,j)=\sum_{\pmb A'}\,\sum_{\pmb B'}\,\sum_{\shuffle}\,C^\epsilon(\pmb B'\shuffle\bar{\pmb A'})\,{\cal A}^{\rm BAS}_n(i,\pmb B'\shuffle\bar{\pmb A'},j|\pmb A,i,\pmb B,j)\,,
~~\label{expan-YM-AB}
\eea
In the above, the summation $\sum_{\shuffle}$ is for all permutations such that the relative order in each of the ordered sets
$\pmb B'$ and $\bar{\pmb A'}$ is kept. For instance, suppose $\pmb B'=\{1,2\}$, $\bar{\pmb A'}=\{3,4\}$, then the summation for shuffles
$\pmb B'\shuffle\bar{\pmb A'}$ is understood as
\bea
\sum_{\shuffle}\,{\cal A}(i,\pmb B'\shuffle\bar{\pmb A'},j)&=&{\cal A}(i,1,2,3,4,j)+{\cal A}(i,1,3,2,4,j)+{\cal A}(i,1,3,4,2,j)\nn
& &+{\cal A}(i,3,4,1,2,j)+{\cal A}(i,3,1,4,2,j)+{\cal A}(i,3,1,2,4,j)\,.
\eea
Two ordered sets $\pmb A'$ and $\pmb B'$ in \eref{expan-YM-AB} are generated from $A$ and $B$ by introducing orderings to elements, respectively, the same as those in \eref{compa-order}. The notation $\bar{\pmb A'}$ means the inverse of $\pmb A'$. For example, if $\pmb A'=\{1,2,3\}$, then $\bar{\pmb A'}=\{3,2,1\}$. Any $\bar{\pmb A'}$ can also be generated from $A$ by giving an ordering to elements. Notice that the expansion \eref{expan-YM-AB}
is just a reformulation of the expansion \eref{expan-YM}. It holds in general, and is independent of the kinematic conditions \eref{kinematic-condi-0-phi3} and \eref{kinematic-condi-0-YM}.

To proceed, we now demand kinematic constraints \eref{kinematic-condi-0-phi3} and \eref{kinematic-condi-0-YM}. The key observation is, under these constraints, any $\pmb r_k$
in the ordered splitting $\{\pmb r_0,\cdots,\pmb r_f\}$ can not include elements in $\pmb A'$ and $\pmb B'$ simultaneously, otherwise the corresponding $F_{\pmb r_k}$ defined in \eref{F0} or \eref{Fl} should vanish. As a consequence, each $C^\epsilon(\pmb B'\shuffle\bar{\pmb A'})$ given in \eref{c-YM} behaves as
\bea
C^\epsilon(\pmb B'\shuffle\bar{\pmb A'})=\sum_{[\bar{\pmb A'},\pmb B']}\,\Big(\prod_{\pmb r_k\subset\bar{\pmb A'}}\,F_{\pmb r_k}\Big)\,\Big(\prod_{\pmb r_k\subset\pmb B'}\,F_{\pmb r_k}\Big)\,,~~\label{c-YM-AB}
\eea
where the summation $\sum_{[\bar{\pmb A'},\pmb B']}$ is for restricted ordered splittings those each $\pmb r_k$ should include elements purely in $\bar{\pmb A'}$ or $\pmb B'$. Meanwhile, if $r^k_1\in\pmb A'$, then the kinematic conditions \eref{kinematic-condi-0-phi3} and \eref{kinematic-condi-0-YM} eliminate all $k_b$ in the combinatorial momentum $\W Z_{r^k_1}$ defined below \eref{Fl}, where $b\in\pmb B'$, and the analogous phenomenon holds when $r^k_1\in\pmb B'$. Therefore, $\prod_{\pmb r_k\subset\bar{\pmb A'}}\,F_{\pmb r_k}$ and $\prod_{\pmb r_k\subset\pmb B'}\,F_{\pmb r_k}$ are independent of shuffles $\pmb B'\shuffle\bar{\pmb A'}$. Consequently, for the given restricted splitting, and given $\pmb A'$ and $\pmb B'$, all ${\cal A}^{\rm BAS}_n(i,\pmb B'\shuffle\bar{\pmb A'},j|\pmb A,i,\pmb B,j)$ share the same pre-factor $\Big(\prod_{\pmb r_k\subset\bar{\pmb A'}}\,F_{\pmb r_k}\Big)\,\Big(\prod_{\pmb r_k\subset\pmb B'}\,F_{\pmb r_k}\Big)$. In other words, we can reorganize \eref{expan-YM-AB} as,
\bea
{\cal A}^{\rm YM}_n(\pmb A,i,\pmb B,j)& &\xrightarrow[]{\eref{kinematic-condi-0-phi3},\eref{kinematic-condi-0-YM}}\nn
& &\sum_{\pmb A'}\,\sum_{\pmb B'}\,\sum_{[\bar{\pmb A'},\pmb B']}\,\Big(\prod_{\pmb r_k\subset\bar{\pmb A'}}\,F_{\pmb r_k}\Big)\,\Big(\prod_{\pmb r_k\subset\pmb B'}\,F_{\pmb r_k}\Big)\,\sum_{\shuffle}\,{\cal A}^{\rm BAS}_n(i,\pmb B'\shuffle\bar{\pmb A'},j|\pmb A,i,\pmb B,j)\,.~~\label{expan-YM-AB2}
\eea

The next step is to convert BAS amplitudes in \eref{expan-YM-AB2} to those vanish as in \eref{zero-BAS}, by applying the well known Kleiss-Kuijf (KK) relation \cite{Kleiss:1988ne}.
The KK relation states
\bea
\sum_{\shuffle}\,{\cal A}_n(i,\pmb B'\shuffle\bar{\pmb A'},j)=(-)^{|\pmb A'|}\,{\cal A}_n(\pmb A',i,\pmb B',j),~~\label{KK}
\eea
which is valid for general ordered amplitudes.
In \eref{KK}, the amplitude at the r.h.s carries the ordering $(\pmb A',i,\pmb B',j)$ which is compatible with the kinematic condition \eref{kinematic-condi-0-phi3},
namely, elements in $\pmb A'$ and $\pmb B'$ are separated by $i$ and $j$.
Substituting \eref{KK} into \eref{expan-YM-AB2}, we get
\bea
{\cal A}^{\rm YM}_n(\pmb A,i,\pmb B,j)& &\xrightarrow[]{\eref{kinematic-condi-0-phi3},\eref{kinematic-condi-0-YM}}\nn
& &\sum_{\pmb A'}\,\sum_{\pmb B'}\,\sum_{[\bar{\pmb A'},\pmb B']}\,\Big(\prod_{\pmb r_k\subset\bar{\pmb A'}}\,F_{\pmb r_k}\Big)\,\Big(\prod_{\pmb r_k\subset\pmb B'}\,F_{\pmb r_k}\Big)\,(-)^{|\pmb A'|}\,{\cal A}^{\rm BAS}_n(\pmb A',i,\pmb B',j|\pmb A,i,\pmb B,j)\,,~~\label{expan-YM-AB3}
\eea
therefore zeros of YM amplitudse ${\cal A}^{\rm YM}_n(\pmb A,i,\pmb B,j)$ in \eref{zero-YM} emergence as the consequence of zeros of BAS amplitudes in \eref{zero-BAS}.

\subsection{NLSM amplitudes}
\label{subsec-NLSM}

Then we discuss hidden zeros of another type of ordered tree-level amplitudes, i.e., the $U(N)$ NLSM ones.
The hidden zeros of tree-level NLSM amplitudes state that \cite{Arkani-Hamed:2023swr}
\bea
{\cal A}^{\rm NLSM}_n(\pmb A,i,\pmb B,j)&\xrightarrow[]{\eref{kinematic-condi-0-phi3}}&0\,,~~\label{zero-NLSM}
\eea
such behavior can be proved similar as in the previous subsection \ref{subsec-YM}.

As in the YM case, we first expand NLSM amplitudes ${\cal A}^{\rm NLSM}_n(\pmb A,i,\pmb B,j)$ to
\bea
{\cal A}^{\rm NLSM}_n(\pmb A,i,\pmb B,j)=\sum_{\pmb A'}\,\sum_{\pmb B'}\,\sum_{\shuffle}\,\hat{C}(\pmb B'\shuffle\bar{\pmb A'})\,{\cal A}^{\rm BAS}_n(i,\pmb B'\shuffle\bar{\pmb A'},j|\pmb A,i,\pmb B,j)\,,
~~\label{expan-NLSM-AB}
\eea
based on the expansion \eref{expan-NLSM}. Again, the above expansion formula is just a reformulation of \eref{expan-BI},
thus holds in general.
The loci \eref{kinematic-condi-0-phi3} reduce the coefficients $\hat C(\pmb B'\shuffle\bar{\pmb A'})$ given in \eref{c-NLSM} to
\bea
\hat C(\pmb B'\shuffle\bar{\pmb A'})=\Big(\prod_{a_\ell\in\bar{\pmb A'}}\,k_{a_\ell}\cdot X^{\bar{\pmb A'}}_{a_\ell}\Big)\,\Big(\prod_{b_\ell\in\pmb B'}\,k_{b_\ell}\cdot X^{\pmb B'}_{b_\ell}\Big)\,\,,~~\label{c-NLSM-AB}
\eea
where $X^{\bar{\pmb A'}}_{a_\ell}=k_i+\W X^{\bar{\pmb A'}}_{a_\ell}$, $X^{\pmb B'}_{b_\ell}=k_i+\W X^{\pmb B'}_{b_\ell}$.
Here $\W X^{\bar{\pmb A'}}_{a_\ell}$ is the summation of momenta carried by external legs at the l.h.s of $a_\ell$ in $\bar{\pmb A'}$,
while $\W X^{\pmb B'}_{b_\ell}$ is the summation of momenta of legs at the l.h.s of $b_\ell$ in $\pmb B'$. Clearly, both $\prod_{a_\ell\in\bar{\pmb A'}}\,k_{a_\ell}\cdot X^{\bar{\pmb A'}}_{a_\ell}$ and $\prod_{b_\ell\in\pmb B'}\,k_{b_\ell}\cdot X^{\pmb B'}_{b_\ell}$
in \eref{c-NLSM-AB} are independent of shuffles $\pmb B'\shuffle\bar{\pmb A'}$. Therefore, the expansion \eref{expan-NLSM-AB} can be reorganized
as
\bea
{\cal A}^{\rm NLSM}_n(\pmb A,i,\pmb B,j)&&\xrightarrow[]{\eref{kinematic-condi-0-phi3}}\nn
& &\sum_{\pmb A'}\,\sum_{\pmb B'}\,\Big(\prod_{a_\ell\in\bar{\pmb A'}}\,k_{a_\ell}\cdot X^{\bar{\pmb A'}}_{a_\ell}\Big)\,\Big(\prod_{b_\ell\in\pmb B'}\,k_{b_\ell}\cdot X^{\pmb B'}_{b_\ell}\Big)\,\sum_{\shuffle}\,{\cal A}^{\rm BAS}_n(i,\pmb B'\shuffle\bar{\pmb A'},j|\pmb A,i,\pmb B,j)\nn
&=&\sum_{\pmb A'}\,\sum_{\pmb B'}\,\Big(\prod_{a_\ell\in\bar{\pmb A'}}\,k_{a_\ell}\cdot X^{\bar{\pmb A'}}_{a_\ell}\Big)\,\Big(\prod_{b_\ell\in\pmb B'}\,k_{b_\ell}\cdot X^{\pmb B'}_{b_\ell}\Big)\,(-)^{|\pmb A'|}\,{\cal A}^{\rm BAS}_n(\pmb A',i,\pmb B',j|\pmb A,i,\pmb B,j)\,,
~~\label{expan-NLSM-AB2}
\eea
where the last equality uses the KK relation \eref{KK}. Consequently, under the constraint \eref{kinematic-condi-0-phi3}, NLSM amplitudes ${\cal A}^{\rm NLSM}_n(\pmb A,i,\pmb B,j)$ vanish due to zeros of BAS amplitudes in \eref{zero-BAS}.

\section{Hidden zeros of unordered amplitudes}
\label{sec-unordered}

This section devote to derive hidden zeros of unordered amplitudes of SG, DBI and GR theories. The main technic is similar as in the previous section, we will convert ordered amplitudes in effective parts of expansions to those carry orderings which are compatible with the kinematic loci \eref{kinematic-condi-0-phi3}, then claim the vanishing of target amplitudes.

As discussed in section \ref{sec-intro}, for unordered amplitudes under consideration in this section, a subtle and interesting problem is the potential divergent propagators, caused by the kinematic condition $s_{ab}=0$. We will study the elimination of such divergences case by case.
Clearly, such divergences correspond to $2$-point channels, thus is automatically absent in SG amplitudes, due to the lacking of cubic interactions.
For DBI amplitudes, the absence of such divergence is indicated by the expansion of target amplitudes to NLSM ones. The most complicated and interesting case is GR, and we will show the detailed mechanism of eliminating such divergences.

\subsection{SG amplitudes}
\label{subsec-SG}

We begin with the simplest unordered amplitudes, of the special situation of Galileon (SG) theory described in \cite{Cachazo:2014xea}.
The same as BAS and NLSM amplitudes, tree-level $n$-point SG amplitudes have hidden zeros \cite{Bartsch:2024amu,Li:2024qfp,Zhang:2024efe}
\bea
A^{\rm SG}_n&\xrightarrow[]{\eref{kinematic-condi-0-phi3}}&0\,.~~\label{zero-SG}
\eea
As mentioned in section \ref{subsec-BAS}, we use $A_n$ to denote full unordered amplitudes with coupling constants striped off.
For the SG case, one need not to worry about the divergences caused by $s_{ab}=0$ in denominators, since the SG theory does not involve any cubic vertex.

A straightforward way to understand the zeros in \eref{zero-SG} is to apply the expansion in \eref{expan-SGtoNLSM}, as well as the manipulation from \eref{expan-NLSM-AB} to \eref{expan-NLSM-AB2}. More explicitly, we first expand $n$-point SG amplitudes $A^{\rm SG}_n$ as
\bea
A^{\rm SG}_n=\sum_{\pmb A}\,\sum_{\pmb B}\,\sum_{\shuffle}\,\hat{C}(\pmb B\shuffle\bar{\pmb A})\,{\cal A}^{\rm NLSM}_n(i,\pmb B\shuffle\bar{\pmb A},j)\,.
~~\label{expan-SG-AB}
\eea
Then, by using the decomposition \eref{c-NLSM-AB} and KK relation \eref{KK}, we get
\bea
A^{\rm SG}_n&\xrightarrow[]{\eref{kinematic-condi-0-phi3}}&\sum_{\pmb A}\,\sum_{\pmb B}\,\Big(\prod_{a_\ell\in\bar{\pmb A}}\,k_{a_\ell}\cdot X^{\bar{\pmb A}}_{a_\ell}\Big)\,\Big(\prod_{b_\ell\in\pmb B}\,k_{b_\ell}\cdot X^{\pmb B}_{b_\ell}\Big)\,\sum_{\shuffle}\,{\cal A}^{\rm NLSM}_n(i,\pmb B\shuffle\bar{\pmb A},j)\nn
&=&\sum_{\pmb A}\,\sum_{\pmb B}\,\Big(\prod_{a_\ell\in\bar{\pmb A}}\,k_{a_\ell}\cdot X^{\bar{\pmb A}}_{a_\ell}\Big)\,\Big(\prod_{b_\ell\in\pmb B}\,k_{b_\ell}\cdot X^{\pmb B}_{b_\ell}\Big)\,(-)^{|\pmb A|}\,{\cal A}^{\rm NLSM}_n(\pmb A,i,\pmb B,j)\,.
~~\label{expan-SG-AB2}
\eea
In the second line, each NLSM amplitude ${\cal A}^{\rm NLSM}_n(\pmb A,i,\pmb B,j)$ carries an ordering compatible with the kinematic condition \eref{kinematic-condi-0-phi3}.
Thus, SG amplitudes vanish according to zeros of NLSM amplitudes in \eref{zero-NLSM}.

\subsection{DBI amplitudes}
\label{subsec-DBI}

Then we study hidden zeros of tree-level DBI amplitudes.
We restrict ourselves to the pure BI sector with the Lagrangian in \eref{Lag-BI}, namely, external particles of each amplitude are solely photons.

The hidden zeros of such tree-level BI amplitudes can be expressed as \cite{Li:2024qfp,Zhang:2024efe}
\bea
A^{\rm BI}_n&\xrightarrow[]{\eref{kinematic-condi-0-phi3},\eref{kinematic-condi-0-YM}}&0\,.~~\label{zero-BI}
\eea
the same as hidden zeros of YM ones in \eref{zero-YM}. Again, the polarization vectors carried by external particles are constrained by the additional condition \eref{kinematic-condi-0-YM}.
From the Lagrangian perspective, the BI amplitudes contain cubic vertices, therefore, the divergent propagators $1/s_{ab}$ should appear when imposing the kinematic condition \eref{kinematic-condi-0-phi3}. However, one can use \eref{expan-BItoNLSM} to expand BI amplitudes $A^{\rm BI}_n$ to NLSM ones, and use the lacking of cubic vertices in NLSM theory, to argue the absence of divergent propagator.

Thus, we first perform the expansion \eref{expan-BItoNLSM}, to express BI amplitudes as
\bea
A^{\rm BI}_n=\sum_{\pmb A}\,\sum_{\pmb B}\,\sum_{\shuffle}\,C^\epsilon(\pmb B\shuffle\bar{\pmb A})\,{\cal A}^{\rm NLSM}_n(i,\pmb B\shuffle\bar{\pmb A},j)\,.~~\label{expan-BI-AB}
\eea
In the above, propagators $1/s_{ab}$ corresponding to $2$-point channels are excluded by NLSM amplitudes ${\cal A}^{\rm NLSM}_n(i,\pmb B\shuffle\bar{\pmb A},j)$.
Then we apply the decomposition \eref{c-YM-AB}, the observation that $\W Z_{r^k_1}$ defined below \eref{Fl} is independent of elements in $\pmb B$ (or $\pmb A$) if $r^k_1\in\pmb A$ (or $r^k_1\in\pmb B$), and the KK relation \eref{KK}, to reorganize \eref{expan-BI-AB} as
\bea
A^{\rm BI}_n& &\xrightarrow[]{\eref{kinematic-condi-0-phi3},\eref{kinematic-condi-0-YM}}\nn
& &\sum_{\pmb A}\,\sum_{\pmb B}\,\sum_{[\bar{\pmb A},\pmb B]}\,\Big(\prod_{\pmb r_k\subset\bar{\pmb A}}\,F_{\pmb r_k}\Big)\,\Big(\prod_{\pmb r_k\subset\pmb B}\,F_{\pmb r_k}\Big)\,\sum_{\shuffle}\,{\cal A}^{\rm NLSM}_n(i,\pmb B\shuffle\bar{\pmb A},j)\nn
&=&\sum_{\pmb A}\,\sum_{\pmb B}\,\sum_{[\bar{\pmb A},\pmb B]}\,\Big(\prod_{\pmb r_k\subset\bar{\pmb A}}\,F_{\pmb r_k}\Big)\,\Big(\prod_{\pmb r_k\subset\pmb B}\,F_{\pmb r_k}\Big)\,(-)^{|\pmb A|}\,{\cal A}^{\rm NLSM}_n(\pmb A,i,\pmb B,j)\,.~~\label{expan-BI-AB2}
\eea
Based on the above expansion formula \eref{expan-BI-AB2}, hidden zeros of BI amplitudes in \eref{zero-BI} are ensured by zeros of NLSM amplitudes in \eref{zero-NLSM}.

\subsection{GR amplitudes}
\label{subsec-GR}

Finally, we consider the most complicated unordered amplitudes, the GR ones. The discussion and conclusion in this subsection is valid for either the pure Einstein gravity, or the extended model that gravity coupled to B-field and dilaton.

As mentioned in section \ref{sec-intro}, GR amplitudes have the similar hidden zeros,
\bea
A^{\rm GR}_n&\xrightarrow[]{\eref{kinematic-condi-0-phi3},\eref{kinematic-condi-0-GR}}&0\,,~~\label{zero-GR}
\eea
with the new constraint on polarizations
\bea
& &\epsilon_a\cdot\epsilon_b=0\,,~~~~\epsilon_a\cdot k_b=0\,,~~~~\epsilon_b\cdot k_a=0\,,\nn
& &\W\epsilon_a\cdot\W\epsilon_b=0\,,~~~~\W\epsilon_a\cdot k_b=0\,,~~~~\W\epsilon_b\cdot k_a=0\,,~~~~{\rm with}~a\in A\,,~b\in B\,.
~~\label{kinematic-condi-0-GR}
\eea
As discussed in section \eref{subsec-expansion}, the polarization tensor of a graviton $\ell$ is decomposed into $\epsilon_\ell$ and $\W\epsilon_\ell$.
The GR amplitudes are more interesting, since the removing of divergences from $1/s_{ab}$ is more subtle. Firstly, GR amplitudes contain cubic interactions, thus we can not exclude these divergences as in subsection \ref{subsec-SG}. Secondly, GR amplitudes can not be expanded to amplitudes without cubic vertices such as NLSM ones, this observation forbids us to remove $1/s_{ab}$ as in subsection \ref{subsec-DBI}.
Therefore, we need to consider divergent terms more carefully.

Similar as before, we use the expansion formula \eref{expan-GR} to expand a tree-level GR amplitude as
\bea
A^{\rm GR}_n=\sum_{\pmb A\,\pmb A'}\,\sum_{\pmb B\,\pmb B'}\,\sum_{\shuffle_1\,\shuffle_2}\,C^\epsilon(\pmb B\shuffle_1\bar{\pmb A})\,C^{\W\epsilon}(\pmb B'\shuffle_2\bar{\pmb A'})\,{\cal A}^{\rm BAS}_n(i,\pmb B\shuffle_1\bar{\pmb A},j|i,\pmb B'\shuffle_2\bar{\pmb A'},j)\,.~~\label{expan-GR-AB}
\eea
To handle the divergences arise from $s_{ab}=0$, we introduce the regularization parameter $\tau$, namely, all vanished Lorentz invariants in
\eref{kinematic-condi-0-phi3} and \eref{kinematic-condi-0-GR} are assumed to be proportional to $\tau$. Now consider a special BAS diagram which contains divergent propagators
\bea
\prod_{t=1}^r\,{1\over s_{a_tb_t}}\,,~~\label{pro-diver}
\eea
with $a_t\in A$ and $b_t\in B$. According to our regularization, the denominator is at the $\tau^r$ order. In the expansion \eref{expan-GR-AB}, the corresponding coefficient of this BAS diagram contains various terms, we classify these terms into three categories.
The first one, terms at the $\tau^0$ order; the second one, terms at the $\tau^q$ order with $0<q<r$; and the third one, terms at the $\tau^p$ order
with $r\leq p$. Clearly, the first category corresponds to restricted ordered splittings those elements in $A$ and $B$ can not enter the same $\pmb r_k$.
\begin{figure}
  \centering
   \includegraphics[width=8.5cm]{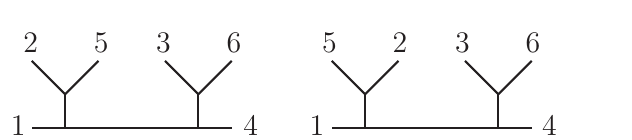}
   \includegraphics[width=8.5cm]{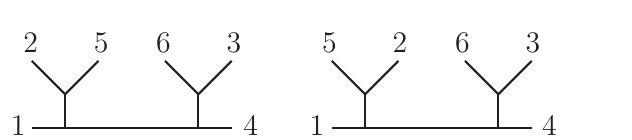} \\
  \caption{Four diagrams which carry propagators ${1\over s_{25}}\,{1\over s_{36}}\,{1\over s_{125}}$. These diagrams are related by swamping $2$ and $5$, or swamping $3$ and $6$.}\label{6p}
\end{figure}

We claim that terms in the second category are canceled when performing the summation in \eref{expan-GR-AB}. To see this, let us begin with an explicit example, the $6$-point diagrams in Fig.\ref{6p}, with $i=1$, $j=4$, $A=\{2,3\}$, $B=\{5,6\}$. All these diagrams contribute the same propagators
\bea
{1\over s_{25}}\,{1\over s_{36}}\,{1\over s_{125}}.~~\label{pro-6p}
\eea
The denominator of \eref{pro-6p} is at the $\tau^2$ order, due to the locus $s_{ab}=0$. Four diagrams in Fig.\ref{6p} correspond to orderings
$(1,2,5,3,6,4)$, $(1,5,2,3,6,4)$, $(1,2,5,6,3,4)$ and $(1,5,2,6,3,4)$, respectively. When $\pmb\sigma_6$ takes one of above four orderings and so does $\pmb\sigma'_6$ (without requiring $\pmb\sigma_6=\pmb\sigma'_6$), then propagators in \eref{pro-6p} are contained in ${\cal A}^{\rm BAS}_6(\pmb\sigma_6|\pmb\sigma'_6)$. Let us consider the corresponding coefficients of these ${\cal A}^{\rm BAS}_6(\pmb\sigma_6|\pmb\sigma'_6)$ in the expansion \eref{expan-GR-AB}. As mentioned previously, terms in coefficients are classified into three categories, and we now focus on the second one.
For the current example, terms in the second category are those at the $\tau^1$ order. We first set the reference ordering to be $6\prec5\prec3\prec2$, and pick up an ordered splitting for the first ordering $(1,2,5,3,6,4)$ (the first diagram in Fig.\ref{6p}), whose contribution is at the $\tau^1$ order under consideration. For instance, we choose the splitting $\{\emptyset,\{3,6\},\{5\},\{2\}\}$. According to rules in \eref{F0} and \eref{Fl}, the kinematic factor $\prod_{k=0}^f\,F_{\pmb r_k}$ of such ordered splitting is given as
\bea
(\epsilon_4\cdot\epsilon_1)\,(\epsilon_6\cdot f_3\cdot k_1)\,(\epsilon_5\cdot k_1)\,(\epsilon_2\cdot k_1)\,.~~\label{6p-term1}
\eea
Then we turn to the second diagram in Fig.\ref{6p}, which corresponds to the ordering $(1,5,2,3,6,4)$. It is directly to see that the same splitting
$\{\emptyset,\{3,6\},\{5\},\{2\}\}$ is also compatible with the new ordering $(1,5,2,3,6)$. Furthermore, this common splitting gives the same term \eref{6p-term1} for the new ordering. On the other hand, the second diagram is related to the first one via a swamping of legs $2$ and $5$ at a vertex, thus the anti-symmetry of structure constants in \eref{Lag-BAS} indicates that a relative $-$ sign will be created. It means, the term obtained by combining \eref{pro-6p} and \eref{6p-term1} enters two BAS amplitudes ${\cal A}^{\rm BAS}_6(1,2,5,3,6,4|\pmb\sigma_6)$ and ${\cal A}^{\rm BAS}_6(1,5,2,3,6,4|\pmb\sigma_6)$ (or, ${\cal A}^{\rm BAS}_6(\pmb\sigma_6|1,2,5,3,6,4)$ and ${\cal A}^{\rm BAS}_6(\pmb\sigma_6|1,5,2,3,6,4)$) with opposite signs. Therefore, when summing over ${\cal A}^{\rm BAS}_6(1,2,5,3,6,4|\pmb\sigma_6)$ and ${\cal A}^{\rm BAS}_6(1,5,2,3,6,4|\pmb\sigma_6)$ with any fixed $\pmb\sigma_6$, such terms cancel each other.

In the above example, we have not used kinematic conditions \eref{kinematic-condi-0-phi3} and \eref{kinematic-condi-0-GR}. To see the role of these kinematic conditions, let us consider another example, the ordered splitting $\{\emptyset,\{5,3,6\},\{2\}\}$. This splitting is again simultaneously compatible with orderings $(1,2,5,3,6,4)$ and $(1,5,2,3,6,4)$ for first and second diagrams in Fig.\ref{6p}. The corresponding kinematic factor $\prod_{k=0}^f\,F_{\pmb r_k}$ determined by this splitting is also the same for two orderings, which is found as
\bea
(\epsilon_6\cdot f_3\cdot f_5\cdot k_1)\,(\epsilon_2\cdot k_1)\,.~~\label{6p-term2}
\eea
In this example, the kinematic constraints are crucial, since without the requirement $\epsilon_2\cdot k_5=0$, we should have $\epsilon_2\cdot k_{15}$
instead of $\epsilon_2\cdot k_1$ for the second ordering $(1,5,2,3,6,4)$. Again, these terms cancel each other when summing over ${\cal A}^{\rm BAS}_6(1,2,5,3,6,4|\pmb\sigma_6)$ and ${\cal A}^{\rm BAS}_6(1,5,2,3,6,4|\pmb\sigma_6)$, due to the same reason.

In appendix \ref{sec-appen2}, we give all cancellations for $\tau^1$ terms of diagrams in Fig.\ref{6p}, for ordered splittings with $\pmb r_0=\emptyset$. As will be seen, the remaining terms are only those at $\tau^2$ and $\tau^3$ orders.

The above discussion about cancellations can be straightforwardly extended to the general case. For any term with the $\tau^r$ order denominator, suppose the corresponding numerator is at the $\tau^q$ order with $0<q<r$, it means at least one pair of legs $(a_t,b_t)$ in \eref{pro-diver}
are not included in the same $\pmb r_k$ in the ordered splitting, since each $\cdots f_{a_t}\cdot f_{b_t}\cdots$ or $\epsilon_{a_t}\cdot f_{b_t}\cdots$
or $\epsilon_{b_t}\cdot f_{a_t}\cdots$ contributes $\tau^1$. If we exchange these separated $a_t$ and $b_t$ in the corresponding ordering, the new ordering is still compatible with the previous ordered splitting. The kinematic factor $\prod_{k=0}^f\,F_{\pmb r_k}$ determined by this splitting is also unaltered, as ensured by kinematic conditions \eref{kinematic-condi-0-phi3} and \eref{kinematic-condi-0-GR}. Meanwhile, swamping $a_t$ and $b_t$ always creates a $-$ sign. Notice that $a_t$ and $b_t$ should be nearby in the ordering, otherwise the propagator $1/s_{a_tb_t}$ can not contribute. Thus, exchanging these two legs will not across other legs, therefore leads to only one $-$ sign. Consequently, when summing over two orderings related via swamping $a_t$ and $b_t$, these terms cancel each other. Such mechanism of cancellation always eliminates all $\tau^q/\tau^r$ terms, due to the following reason. For a given set \eref{pro-diver} of divergent propagators, and a given ordered splitting compatible with these propagators, suppose $u$ pairs of $(a_t,b_t)$ have not been absorbed into the same $\pmb r_k$, then their are $2^u$ corresponding orderings linked by swamping these $(a_t,b_t)$. These $2^u$ orderings share the given ordered splitting, and the same kinematic factor $\prod_{k=0}^f\,F_{\pmb r_k}$. We can choose a fiducial pair $(\dot{a}_{t},\dot{b}_{t})$, and group these orderings into $2^{u-1}$ pairs, in each pair two orderings are related by flipping two fiducial legs $\dot{a}_{t}$ and $\dot{b}_{t}$. Then the divergent $\tau^q/\tau^r$ contributions from two orderings in the same pair cancel each other when doing the summation in \eref{expan-GR-AB}. The above scheme is also demonstrated in appendix \ref{sec-appen2}. We have verified such cancellations up to $8$-point.

To proceed, we turn back to the expansion \eref{expan-GR-AB}, and reform it as
\bea
A^{\rm GR}_n& &\xrightarrow[]{\eref{kinematic-condi-0-phi3},\eref{kinematic-condi-0-GR}}\nn
& &\sum_{\pmb A\,\pmb A'}\,\sum_{\pmb B\,\pmb B'}\,\sum_{\shuffle_1\,\shuffle_2}\,\sum_{{\rm splitting}^L}\,\sum_{{\rm splitting}^R}\,\Big(\prod_{\pmb r_k}\,F_{\pmb r_k}\Big)\,\Big(\prod_{\pmb r_k}\,\W F_{\pmb r_k}\Big)\,{\cal A}^{\rm BAS}_n(i,\pmb B\shuffle_1\bar{\pmb A},j|i,\pmb B'\shuffle_2\bar{\pmb A'},j)\,,~~\label{expan-GR-AB2}
\eea
where the summation $\sum_{{\rm splitting}^L}$ is for ordered splittings of left orderings $(i,\pmb B\shuffle_1\bar{\pmb A},j)$, while $\sum_{{\rm splitting}^R}$ is for right orderings $(i,\pmb B'\shuffle_1\bar{\pmb A'},j)$. The notation $\W F_{\pmb r_k}$ is introduced to emphasize that all polarization vectors contribute to $\W F_{\pmb r_k}$ are in $\{\W\epsilon_\ell\}$.
We divide the summation $\sum_{{\rm splitting}^L}\sum_{{\rm splitting}^R}$ into
\bea
\sum_{{\rm splitting}^L}\sum_{{\rm splitting}^R}=\sum_{[\bar{\pmb A},\pmb B]|{\rm all}}+\sum_{{\rm all}|[\bar{\pmb A'},\pmb B']}+\sum_{\overline{[\bar{\pmb A},\pmb B]}|\overline{[\bar{\pmb A'},\pmb B']}}-\sum_{[\bar{\pmb A},\pmb B]|[\bar{\pmb A'},\pmb B']}\,.~~\label{divi-sum}
\eea
In the above, the summation $\sum_{[\bar{\pmb A},\pmb B]|{\rm all}}$ means summing over restricted ordered splittings those elements from $A$ and $B$ are separated into different $\pmb r_k$, for left orderings $(i,\pmb B\shuffle_1\bar{\pmb A},j)$ in \eref{expan-GR-AB2}, and summing over all ordered splittings for right orderings $(i,\pmb B'\shuffle_1\bar{\pmb A'},j)$.
The second summation $\sum_{{\rm all}|[\bar{\pmb A'},\pmb B']}$ is understood by exchanging left and right orderings in the first one.
The third summation is among un-restricted ordered splittings for both left and right orderings, namely, ordered splittings those at least one pair $(a,b)$ with $a\in A$ and $b\in B$ enters the same $\pmb r_k$.
The last summation in \eref{divi-sum} is for restricted ordered splittings for both left and right orderings, it is introduced to eliminate the over counting in first two summations.

Now the hidden zeros of GR amplitudes can be observed as follows. For the first summation in \eref{divi-sum}, we can apply the expansion \eref{expan-YM}, the decomposition \eref{c-YM-AB}, as well as the KK relation \eref{KK},
to organize this part as
\bea
& &\sum_{\pmb A\,\pmb A'}\,\sum_{\pmb B\,\pmb B'}\,\sum_{\shuffle_1\,\shuffle_2}\,\sum_{[\bar{\pmb A},\pmb B]|{\rm all}}\,\Big(\prod_{\pmb r_k}\,F_{\pmb r_k}\Big)\,\Big(\prod_{\pmb r_k}\,\W F_{\pmb r_k}\Big)\,{\cal A}^{\rm BAS}_n(i,\pmb B\shuffle_1\bar{\pmb A},j|i,\pmb B'\shuffle_2\bar{\pmb A'},j)\nn
&&~~~~~~~~\xrightarrow[]{\eref{kinematic-condi-0-phi3},\eref{kinematic-condi-0-GR}}\,\sum_{\pmb A}\,\sum_{\pmb B}\,\sum_{\shuffle}\,\sum_{[\bar{\pmb A},\pmb B]}\,\Big(\prod_{\pmb r_k}\,F_{\pmb r_k}\Big)\,{\cal A}^{\rm YM}_n(i,\pmb B\shuffle\bar{\pmb A},j)\nn
&&~~~~~~~~~~~~~~~~~~=\sum_{\pmb A}\,\sum_{\pmb B}\,\sum_{[\bar{\pmb A},\pmb B]}\,\Big(\prod_{\pmb r_k\subset\bar{\pmb A}}\, F_{\pmb r_k}\Big)\,\Big(\prod_{\pmb r_k\subset\pmb B}\, F_{\pmb r_k}\Big)\,\sum_{\shuffle}\,{\cal A}^{\rm YM}_n(i,\pmb B\shuffle\bar{\pmb A},j)\nn
&&~~~~~~~~~~~~~~~~~~=\sum_{\pmb A}\,\sum_{\pmb B}\,\sum_{[\bar{\pmb A},\pmb B]}\,\Big(\prod_{\pmb r_k\subset\bar{\pmb A}}\, F_{\pmb r_k}\Big)\,\Big(\prod_{\pmb r_k\subset\pmb B}\, F_{\pmb r_k}\Big)\,(-)^{|\pmb A|}\,{\cal A}^{\rm YM}_n(\pmb A,i,\pmb B,j)\,.~~\label{expan-GR-AB3}
\eea
The last line of above formula contains no divergent propagator, and vanishes due to the zeros of YM amplitudes in \eref{zero-YM}.
The parallel manipulation holds for the second summation in \eref{divi-sum}, and the vanishing of this part is guaranteed by the same reason. In the third summation, for a given $\tau^r$ denominator, the effective components in $\prod_{\pmb r_k}\,F_{\pmb r_k}$ and $\prod_{\pmb r_k}\,\W F_{\pmb r_k}$ are those at $\tau^p$ and $\tau^{\W p}$ orders satisfying $r\leq p$, $r\leq \W p$, based on cancellations of $\tau^q$ and $\tau^{\W q}$ terms. Thus, this part vanishes in the limit $\tau\to0$, since $p+\W p>r$.
For the last summation, we can use the technic similar as in \eref{expan-GR-AB3}, to obtain
\bea
& &\sum_{\pmb A\,\pmb A'}\,\sum_{\pmb B\,\pmb B'}\,\sum_{\shuffle_1\,\shuffle_2}\,\sum_{[\bar{\pmb A},\pmb B]|[\bar{\pmb A'},\pmb B']}\,\Big(\prod_{\pmb r_k}\,F_{\pmb r_k}\Big)\,\Big(\prod_{\pmb r_k}\,\W F_{\pmb r_k}\Big)\,{\cal A}^{\rm BAS}_n(i,\pmb B\shuffle_1\bar{\pmb A},j|i,\pmb B'\shuffle_2\bar{\pmb A'},j)\nn
&&~~~~~~~~\xrightarrow[]{\eref{kinematic-condi-0-phi3},\eref{kinematic-condi-0-GR}}\,\sum_{\pmb A\,\pmb A'}\,\sum_{\pmb B\,\pmb B'}\,\sum_{[\bar{\pmb A},\pmb B]|[\bar{\pmb A'},\pmb B']}\,\Big(\prod_{\pmb r_k\subset\bar{\pmb A}}\, F_{\pmb r_k}\Big)\,\Big(\prod_{\pmb r_k\subset\pmb B}\, F_{\pmb r_k}\Big)\,\Big(\prod_{\pmb r_k\subset\bar{\pmb A'}}\, \W F_{\pmb r_k}\Big)\,\Big(\prod_{\pmb r_k\subset\pmb B'}\,\W F_{\pmb r_k}\Big)\nn
&&~~~~~~~~~~~~~~~~~~~~~~~~~~~~~~~~~~~~~~~~~~~~~~~~\sum_{\shuffle_1\,\shuffle_2}\,{\cal A}^{\rm BAS}_n(i,\pmb B\shuffle_1\bar{\pmb A},j|i,\pmb B'\shuffle_2\bar{\pmb A'},j)\nn
&&~~~~~~~~~~~~~~~~~~=\sum_{\pmb A\,\pmb A'}\,\sum_{\pmb B\,\pmb B'}\,\sum_{[\bar{\pmb A},\pmb B]|[\bar{\pmb A'},\pmb B']}\,\Big(\prod_{\pmb r_k\subset\bar{\pmb A}}\, F_{\pmb r_k}\Big)\,\Big(\prod_{\pmb r_k\subset\pmb B}\, F_{\pmb r_k}\Big)\,\Big(\prod_{\pmb r_k\subset\bar{\pmb A'}}\, \W F_{\pmb r_k}\Big)\,\Big(\prod_{\pmb r_k\subset\pmb B'}\,\W F_{\pmb r_k}\Big)\nn
&&~~~~~~~~~~~~~~~~~~~~~~~~~~~~~~~~~~~~~~~~~~~~~~~~~~{\cal A}^{\rm BAS}_n(\pmb A,i,\pmb B,j|\pmb A',i,\pmb B',j)\,,~~\label{expan-GR-AB4}
\eea
therefore the vanishing of this part is governed by zeros of BAS amplitudes in \eref{zero-BAS}.

We have seen that, the absence of divergences in the GR case is much more non-trivial than in SG and DBI cases. Before ending this subsection, we briefly summarize the process of eliminating such divergences. Firstly, the divergent $\tau^q/\tau^r$ terms are canceled. Secondly, the remaining effective parts in the third summation in \eref{divi-sum}, which carries divergent propagators, vanishes since the coefficients are at higher orders of $\tau$. Finally, the remaining divergent propagators are removed via the KK relation, as shown in \eref{expan-GR-AB3} and \eref{expan-GR-AB4}.

\section{Summary}
\label{sec-summary}

In this note, we have re-derived hidden zeros of tree-level YM, NLSM, SG, DBI and GR amplitudes, by utilizing universal expansions. According to our method, zeros of amplitudes of above theories are ultimately attributed to zeros of BAS amplitudes, while such zeros of BAS amplitudes can be directly proved. The potential divergences caused by the kinematic condition $s_{ab}=0$, which may enter the unordered amplitudes, are discussed in detail. We have shown that, such divergences can always be excluded for unordered amplitudes under consideration. For SG and DBI cases, they are straightforwardly forbidden. For GR amplitudes, the elimination of divergences is much more non-trivial, but all these divergences are cancelled or removed at the end.

Although we restrict the discussions to amplitudes with pure external states, namely each amplitude contains only one type of external particles,
the method used in this note can be naturally extended to mixed amplitudes those different types of particles coupled together. For instance, by using the expansions of YM$\oplus$BAS and NLSM$\oplus$BAS amplitudes given in \cite{Zhou:2019mbe} and \cite{Zhou:2024qjh} respectively, one can obtain hidden zeros of these amplitudes, through the technic extremely similar as in sections \ref{subsec-YM} and \ref{subsec-NLSM}. Thus, the hidden zeros of tree-level amplitudes can be directly extended to a broader range of theories.

In this note, we only studied the newly discovered hidden zeros, without discussing the related novel factorizations/splittings. There are some technical obstacles for applying the method in this note to consider factorizations/splittings. For example, in $2$-splits studied in \cite{Cao:2024gln,Cao:2024qpp,Arkani-Hamed:2024fyd}, each amplitude is split into two amputated currents. Suppose we use the method in this note to investigate such $2$-splits, it is natural to expect that the resulting currents are also expressed as expansions to appropriate basis. However, expansions reviewed in section \ref{subsec-expansion} are for on-shell amplitudes. For each amputated current with one off-shell external leg, the analogous expansion formula is not clear. We leave the generalization of our method to factorizations as a future work.

\section*{Acknowledgments}

The authors would thank Yong Zhang, Jin Dong and Liang Zhang for valuable discussions. We also thank Prof. Bo Feng and Liang Zhang for cooperation on related topics. This work is supported by NSFC under Grant No. 11805163.


\appendix

\section{Proof of hidden zeros of tree-level BAS amplitudes}
\label{sec-appen}

In this section, we give a proof of hidden zeros of BAS amplitudes in \eref{zero-BAS}, by solely utilizing Feynman rules. This proof was proposed in \cite{Zhou:2024ddy}.

We start by considering the special case that two orderings ${\pmb\sigma}_n$ and ${\pmb\sigma}'_n$ are equivalent to each other. Such amplitudes are called the ${\rm Tr}(\phi^3)$ ones.
The zeros of an $n$-point ${\rm Tr}(\phi^3)$ amplitude is given as
\bea
{\cal A}^{{\rm Tr}(\phi^3)}_n(\pmb A,i,\pmb B,j)&\xrightarrow[]{\eref{kinematic-condi-0-phi3}}&0\,,~~\label{zero-phi3}
\eea
where $\pmb A$ and $\pmb B$ are two ordered sets among external legs, satisfying
\bea
{\pmb \sigma}_n={\pmb \sigma}'_n=\pmb A,i,\pmb B,j\,,
\eea
up to cyclic permutations.

\begin{figure}
  \centering
  \includegraphics[width=11cm]{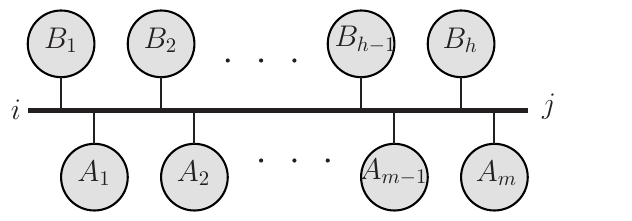} \\
  \caption{General feature of diagrams of ordered ${\rm Tr}(\phi^3)$ amplitudes. One can always find a line $L_{ij}$ which connect two external legs $i$ and $j$, and think the remaining part of each diagram as planting blocks to the line $L_{ij}$. }\label{npphi3}
\end{figure}

The interesting behavior in \eref{zero-phi3} can be interpreted as follows.
For each Feynman diagram contributes to ${\cal A}^{{\rm Tr}(\phi^3)}_n(\pmb A,i,\pmb B,j)$,
one can always find a line $L_{i,j}$ which connects two legs $i$ and $j$, and think the diagram as planting blocks onto
$L_{i,j}$, as can be seen in Fig.\ref{npphi3}. In Fig.\ref{npphi3}, each external leg $a\in\pmb A$ belongs to one of block
$A_c$ with $c\in\{1,\cdots,m\}$, while each $b\in\pmb B$ belongs to one of block $B_d$ with $d\in\{1,\cdots,h\}$.
Based on the above picture, the full amplitude can be expressed as
\bea
{\cal A}^{{\rm Tr}(\phi^3)}_n(1,\cdots,n)&=&\sum_{\rm division}\,\Big(\prod_{r=1}^m\,{\cal J}_{A_r}\Big)\,
\Big(\prod_{r=1}^{h}\,{\cal J}_{B_r}\Big)\,\Big(\sum_{\Gamma^{i,j}_{m,h}}\,\prod_{t=1}^{m+h-1}{1\over{\cal D}^{i,j}_t}\Big)\,,~~\label{np-for0}
\eea
where $1/{\cal D}^{i,j}_t$ stand for propagators along $L_{i,j}$, ${\cal J}_{A_r}$ and ${\cal J}_{B_r}$ serve as Berends-Giele currents \cite{Berends:1987me} from blocks. The summation $\sum_{\rm division}$ is for
divisions which separate external legs in $\pmb A$ and $\pmb B$ into different blocks, while the summation $\sum_{\Gamma^{i,j}_{m,h}}$
is for diagrams with given $A_c$ and $B_c$, respecting to the ordering among external legs.
The key observation is
\bea
\sum_{\Gamma^{i,j}_{m,h}}\,\prod_{t=1}^{m+h-1}{1\over{\cal D}^{i,j}_t}
&=&s_{iA_1\cdots A_m B_1\cdots B_h}\,\Big(\prod_{c=1}^{m}\,{1\over s_{iA_1\cdots A_c}}\Big)\,\Big(\prod_{d=1}^{h}\,{1\over s_{iB_1\cdots B_d}}\Big)\nn
&=&k_j^2\,\Big(\prod_{c=1}^{m}\,{1\over s_{iA_1\cdots A_c}}\Big)\,\Big(\prod_{d=1}^{h}\,{1\over s_{iB_1\cdots B_d}}\Big)\,,~~\label{for0-step2}
\eea
thus the amplitude \eref{np-for0} vanishes due to the on-shell condition $k_j^2=0$.
The simplest example of \eref{for0-step2} is $m=h=1$, one can directly verify that
\bea
{1\over s_{iA_1}}+{1\over s_{iB_1}}={s_{iA_1B_1}\over s_{iA_1}\,s_{iB_1}}\,,
\eea
since the locus \eref{kinematic-condi-0-phi3} indicates $s_{iA_1}+s_{iB_1}=s_{iA_1B_1}$. The general behavior \eref{for0-step2}
can be proved recursively, as shown in \cite{Zhou:2024ddy}.

\begin{figure}
  \centering
  \includegraphics[width=13cm]{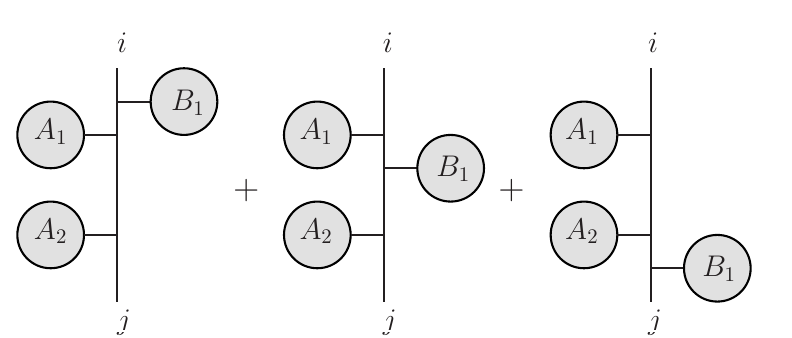} \\
  \caption{Summing over diagrams with $m=2$, $h=1$.}\label{fac-exam}
\end{figure}

As another example, let us consider the division $m=2$, $h=1$, i.e., three blocks
$A_1$, $A_2$ and $B_1$ are planted to $L_{i,j}$. The corresponding Feynman diagrams, which are compatible with the ordering among external legs,
are given in Fig.\ref{fac-exam}. Such diagrams contribute
\bea
C_{A_1A_2B_1}={\cal J}_{A_1}\,{\cal J}_{A_2}\,{\cal J}_{B_1}\,\Big({1\over s_{iB_1}}{1\over s_{iA_1B_1}}+{1\over s_{iA_1}}{1\over s_{iA_1B_1}}+{1\over s_{iA_1}}{1\over s_{iA_1A_2}}\Big)\,.
\eea
For this special case, we have
\bea
{1\over s_{iB_1}}{1\over s_{iA_1B_1}}+{1\over s_{iA_1}}{1\over s_{iA_1B_1}}+{1\over s_{iA_1}}{1\over s_{iA_1A_2}}
&=&{1\over s_{iA_1}}{1\over s_{iB_1}}+{1\over s_{iA_1}}{1\over s_{iA_1A_2}}\nn
&=&s_{iA_1A_2B_1}\,\Big({1\over s_{iA_1}}{1\over s_{iA_1A_2}}\Big)\,{1\over s_{iB_1}}\,,
\eea
where $s_{iA_1}+s_{iB_1}=s_{iA_1B_1}$ and $s_{iA_1A_2}+s_{iB_1}=s_{iA_1A_2B_1}$ have been used.
Therefore, $C_{A_1A_2B_1}$ vanishes due to $s_{iA_1A_2B_1}=k_j^2=0$.

Now we turn to BAS amplitudes with different orderings ${\pmb\sigma}_n$ and ${\pmb\sigma}'_n$. For general ${\pmb\sigma}_n$ and ${\pmb\sigma}'_n$, the zeros of ${\rm Tr}(\phi^3)$
amplitudes can not be extended to the BAS case. However, if two orderings ${\pmb\sigma}_n$ and ${\pmb\sigma}'_n$ satisfy
the requirement \eref{compa-order}, then the argument for the
${\rm Tr}(\phi^3)$ case still holds. The difference between $\pmb A'$ (or $\pmb B'$) and $\pmb A$ (or $\pmb B$) eliminates various diagrams, but the contribution from remaining diagrams also takes the form \eref{for0-step2}, although the summation $\sum_{\rm division}$ is reduced to a smaller set of divisions. Thus we can conclude the zeros in \eref{zero-BAS}.

\section{Cancellations in the $6$-point example in section \ref{subsec-GR}}
\label{sec-appen2}

In this section, we show all cancellations of $\tau^q/\tau^r$ terms correspond to Feynman diagrams in Fig.\ref{6p}, for ordered splittings satisfying $\pmb r_0=\emptyset$. Four diagrams can be labeled by four associated orderings, $(1,2,5,3,6,4)$, $(1,5,2,3,6,4)$, $(1,2,5,6,3,4)$, $(1,5,2,6,3,4)$.
As in section \ref{subsec-GR}, we choose the reference ordering to be $6\prec5\prec3\prec2$. Then all un-restricted ordered splittings for the above four orderings, can be found as in Table.\ref{order1}, Table.\ref{order2}, Table.\ref{order3}, Table.\ref{order4}, respectively. As can be seen, in these tables, the ordered splittings as well as the corresponding kinematic factors, labeled by the same letter but different subscripts, are exactly the same. For example, $b_1$, $b_2$, $b_3$ and $b_4$ label the same ordered splitting and kinematic factor.

\begin{table}[!h]
\begin{center}
\begin{tabular}{|c|c|c|c|}
\hline
Label& Ordered splitting  & Kinematic factor & Order\\
\hline
$a_1$ & $\{\emptyset,\{3,6\},\{5\},\{2\}\}$ & $(\epsilon_4\cdot \epsilon_1)\,(\epsilon_6\cdot f_3\cdot k_1)\,(\epsilon_5\cdot k_1)\,(\epsilon_2\cdot k_1)$ & $\tau^1$ \\
$b_1$ & $\{\emptyset,\{2,6\},\{5\},\{3\}\}$ & $(\epsilon_4\cdot \epsilon_1)\,(\epsilon_6\cdot f_2\cdot k_1)\,(\epsilon_5\cdot k_1)\,(\epsilon_3\cdot k_{12})$ &$\tau^1$ \\
$c_1$ & $\{\emptyset,\{2,3,6\},\{5\}\}$ & $(\epsilon_4\cdot \epsilon_1)\,(\epsilon_6\cdot f_3\cdot f_2\cdot k_1)\,(\epsilon_5\cdot k_1)$ & $\tau^1$ \\
$d_1$ & $\{\emptyset,\{5,3,6\},\{2\}\}$ & $(\epsilon_4\cdot \epsilon_1)\,(\epsilon_6\cdot f_3\cdot f_5\cdot k_1)\,(\epsilon_2\cdot k_1)$ & $\tau^2$ \\
$e_1$ & $\{\emptyset,\{6\},\{2,5\},\{3\}\}$ & $(\epsilon_4\cdot \epsilon_1)\,(\epsilon_6\cdot k_1)\,(\epsilon_5\cdot f_2\cdot k_1)\,(\epsilon_3\cdot k_{12})$ & $\tau^1$\\
$f_1$ &  $\{\emptyset,\{2,5,6\},\{3\}\}$ & $(\epsilon_4\cdot \epsilon_1)\,(\epsilon_6\cdot f_5\cdot f_2\cdot k_1)\,(\epsilon_3\cdot k_{12})$ & $\tau^1$\\
$h_1$ & $\{\emptyset,\{3,6\},\{2,5\}\}$ & $(\epsilon_4\cdot \epsilon_1)\,(\epsilon_6\cdot f_3\cdot k_1)\,(\epsilon_5\cdot f_2\cdot k_1)$ & $\tau^2$\\
$i_1$ & $\{\emptyset,\{2,3,5,6\}\}$ & $(\epsilon_4\cdot \epsilon_1)\,(\epsilon_6\cdot f_5\cdot f_3\cdot f_2\cdot k_1)$ & $\tau^3$\\
\hline
\end{tabular}
\end{center}
\caption{\label{order1}Ordering $(1,2,5,3,6,4)$}
\end{table}
\begin{table}[!h]
\begin{center}
\begin{tabular}{|c|c|c|c|}
\hline
Label& Ordered splitting  & Kinematic factor & Order\\
\hline
$a_2$ & $\{\emptyset,\{3,6\},\{5\},\{2\}\}$ & $(\epsilon_4\cdot \epsilon_1)\,(\epsilon_6\cdot f_3\cdot k_1)\,(\epsilon_5\cdot k_1)\,(\epsilon_2\cdot k_1)$ & $\tau^1$ \\
$b_2$ & $\{\emptyset,\{2,6\},\{5\},\{3\}\}$ & $(\epsilon_4\cdot \epsilon_1)\,(\epsilon_6\cdot f_2\cdot k_1)\,(\epsilon_5\cdot k_1)\,(\epsilon_3\cdot k_{12})$ &$\tau^1$ \\
$c_2$ & $\{\emptyset,\{2,3,6\},\{5\}\}$ & $(\epsilon_4\cdot \epsilon_1)\,(\epsilon_6\cdot f_3\cdot f_2\cdot k_1)\,(\epsilon_5\cdot k_1)$ & $\tau^1$ \\
$d_2$ & $\{\emptyset,\{5,3,6\},\{2\}\}$ & $(\epsilon_4\cdot \epsilon_1)\,(\epsilon_6\cdot f_3\cdot f_5\cdot k_1)\,(\epsilon_2\cdot k_1)$ & $\tau^2$ \\
$g_2$ & $\{\emptyset,\{5,2,6\},\{3\}\}$ & $(\epsilon_4\cdot \epsilon_1)\,(\epsilon_6\cdot f_2\cdot f_5\cdot k_1)\,(\epsilon_3\cdot k_{12})$ & $\tau^2$\\
$j_2$ &  $\{\emptyset,\{5,2,3,6\}\}$ & $(\epsilon_4\cdot \epsilon_1)\,(\epsilon_6\cdot f_3\cdot f_2\cdot f_5\cdot k_1)$ & $\tau^2$\\
\hline
\end{tabular}
\end{center}
\caption{\label{order2}Ordering $(1,5,2,3,6,4)$}
\end{table}
\begin{table}[!h]
\begin{center}
\begin{tabular}{|c|c|c|c|}
\hline
Label& Ordered splitting  & Kinematic factor & Order\\
\hline
$b_3$ & $\{\emptyset,\{2,6\},\{5\},\{3\}\}$ & $(\epsilon_4\cdot \epsilon_1)\,(\epsilon_6\cdot f_2\cdot k_1)\,(\epsilon_5\cdot k_1)\,(\epsilon_3\cdot k_{12})$ &$\tau^1$ \\
$e_3$ & $\{\emptyset,\{6\},\{2,5\},\{3\}\}$ & $(\epsilon_4\cdot \epsilon_1)\,(\epsilon_6\cdot k_1)\,(\epsilon_5\cdot f_2\cdot k_1)\,(\epsilon_3\cdot k_{12})$ & $\tau^1$\\
$f_3$ &  $\{\emptyset,\{2,5,6\},\{3\}\}$ & $(\epsilon_4\cdot \epsilon_1)\,(\epsilon_6\cdot f_5\cdot f_2\cdot k_1)\,(\epsilon_3\cdot k_{12})$ & $\tau^1$\\
\hline
\end{tabular}
\end{center}
\caption{\label{order3}Ordering $(1,2,5,6,3,4)$}
\end{table}
\begin{table}[!h]
\begin{center}
\begin{tabular}{|c|c|c|c|}
\hline
Label& Ordered splitting  & Kinematic factor & Order\\
\hline
$b_4$ & $\{\emptyset,\{2,6\},\{5\},\{3\}\}$ & $(\epsilon_4\cdot \epsilon_1)\,(\epsilon_6\cdot f_2\cdot k_1)\,(\epsilon_5\cdot k_1)\,(\epsilon_3\cdot k_{12})$ &$\tau^1$ \\
$g_4$ & $\{\emptyset,\{5,2,6\},\{3\}\}$ & $(\epsilon_4\cdot \epsilon_1)\,(\epsilon_6\cdot f_2\cdot f_5\cdot k_1)\,(\epsilon_3\cdot k_{12})$ & $\tau^2$\\
\hline
\end{tabular}
\end{center}
\caption{\label{order4}Ordering $(1,5,2,6,3,4)$}
\end{table}

We can categorize un-restricted splittings in these tables into three types
\bea
& &{\rm Type}-1:~\{a_i,b_i,c_i,d_i\}\,,\nn
& &{\rm Type}-2:~\{e_i,f_i,g_i\}\,,\nn
& &{\rm Type}-3:~\{h_i,i_i,j_i\}\,.
\eea
The above classification is based on the following observation. For propagators in \eref{pro-6p}, their are two pairs $(a_t,b_t)$ which are $(2,5)$ and $(3,6)$, according to the label of divergent propagators in \eref{pro-diver}. The Type-$1$ splittings are those legs $(2,5)$ are separated into different $\pmb r_k$, the Type-$2$ splittings are those legs in $(3,6)$ are separated into different $\pmb r_k$, and the Type-$3$ splittings are those non of $(2,5)$ and $(3,6)$ are separated into different $\pmb r_k$. Notice that the splitting $b_i$ is special.
It can also be understood as in Type-$2$, since either of $(2,5)$ and $(3,6)$ are separated into different $\pmb r_k$.

Four orderings $(1,2,5,3,6,4)$, $(1,5,2,3,6,4)$, $(1,2,5,6,3,4)$ and $(1,5,2,6,3,4)$ are related by swamping legs $2$ and $5$, or swamping legs $3$ and $6$. Since flipping two legs always creates a $-$ sign, we immediately observe the following cancellations.
For Type-$1$ splittings, the cancellations can be grouped into $(a_1,a_2)$, $(b_1,b_2)$, $(c_1,c_2)$, $(d_1,d_2)$ and $(b_3,b_4)$. When summing over different orderings, two kinematic factors in any one of above four pairs cancel each other. For Type-$2$ splittings, the cancellations
are grouped as $(e_1,e_3)$, $(f_1,f_3)$, $(g_2,g_4)$. For Type-$3$ splittings, no cancellation can be found. Therefore, the remaining kinematic factors are those in Type-$3$, labeled as $h_1$, $i_1$ and $j_2$. As can be seen in Table.\ref{order1} and Table.\ref{order2}, these terms are at $\tau^2$ and $\tau^3$ orders. In other words, all terms at the $\tau^1$ order have been eliminated.

In the above, the cancellations of terms from $b_i$ are organized as $(b_1,b_2)$, $(b_3,b_4)$. Such arrangement can be understood as follows. We choose the fiducial pair of legs as $(2,5)$, and group four orderings into two pairs which are $((1,2,5,3,6,4),(1,5,2,3,6,4))$ and $((1,2,5,6,3,4),(1,5,2,6,3,4))$. In each pair, two orderings are related by exchanging legs $2$ and $5$. The cancellations $(b_1,b_2)$ are $(b_3,b_4)$
are for two pairs of orderings respectively, caused by the emergent $-$ sign when swamping legs $2$ and $5$. One can also choose the fiducial pair as $(3,6)$. For this new choice, the cancellations should be organized as $(b_1,b_3)$, $(b_2,b_4)$, since they are caused by swamping legs $3$ and $6$.
For the current simple example, the above scheme is not efficient, since cancellations can be directly observed. However, we can use the above scheme to ensure the general cancellations of $\tau^q/\tau^r$ terms, as discussed in section \ref{subsec-GR}. 

\bibliographystyle{JHEP}

\bibliography{reference}

\providecommand{\href}[2]{#2}\begingroup\raggedright\begin{thebibliography}{10}

\bibitem{Bern:2008qj}
Z.~Bern, J.~J.~M. Carrasco, and H.~Johansson, {\it {New Relations for
  Gauge-Theory Amplitudes}},  {\em Phys. Rev. D} {\bf 78} (2008) 085011,
  [\href{http://arxiv.org/abs/0805.3993}{{\tt arXiv:0805.3993}}].

\bibitem{Bern:2010ue}
Z.~Bern, J.~J.~M. Carrasco, and H.~Johansson, {\it {Perturbative Quantum
  Gravity as a Double Copy of Gauge Theory}},  {\em Phys. Rev. Lett.} {\bf 105}
  (2010) 061602, [\href{http://arxiv.org/abs/1004.0476}{{\tt
  arXiv:1004.0476}}].

\bibitem{Bern:2019prr}
Z.~Bern, J.~J. Carrasco, M.~Chiodaroli, H.~Johansson, and R.~Roiban, {\it {The
  duality between color and kinematics and its applications}},  {\em J. Phys.
  A} {\bf 57} (2024), no.~33 333002,
  [\href{http://arxiv.org/abs/1909.01358}{{\tt arXiv:1909.01358}}].

\bibitem{Cachazo:2013hca}
F.~Cachazo, S.~He, and E.~Y. Yuan, {\it {Scattering of Massless Particles in
  Arbitrary Dimensions}},  {\em Phys. Rev. Lett.} {\bf 113} (2014), no.~17
  171601, [\href{http://arxiv.org/abs/1307.2199}{{\tt arXiv:1307.2199}}].

\bibitem{Cachazo:2013iea}
F.~Cachazo, S.~He, and E.~Y. Yuan, {\it {Scattering of Massless Particles:
  Scalars, Gluons and Gravitons}},  {\em JHEP} {\bf 07} (2014) 033,
  [\href{http://arxiv.org/abs/1309.0885}{{\tt arXiv:1309.0885}}].

\bibitem{Cachazo:2014nsa}
F.~Cachazo, S.~He, and E.~Y. Yuan, {\it {Einstein-Yang-Mills Scattering
  Amplitudes From Scattering Equations}},  {\em JHEP} {\bf 01} (2015) 121,
  [\href{http://arxiv.org/abs/1409.8256}{{\tt arXiv:1409.8256}}].

\bibitem{Cachazo:2014xea}
F.~Cachazo, S.~He, and E.~Y. Yuan, {\it {Scattering Equations and Matrices:
  From Einstein To Yang-Mills, DBI and NLSM}},  {\em JHEP} {\bf 07} (2015) 149,
  [\href{http://arxiv.org/abs/1412.3479}{{\tt arXiv:1412.3479}}].

\bibitem{Arkani-Hamed:2017mur}
N.~Arkani-Hamed, Y.~Bai, S.~He, and G.~Yan, {\it {Scattering Forms and the
  Positive Geometry of Kinematics, Color and the Worldsheet}},  {\em JHEP} {\bf
  05} (2018) 096, [\href{http://arxiv.org/abs/1711.09102}{{\tt
  arXiv:1711.09102}}].

\bibitem{Arkani-Hamed:2023lbd}
N.~Arkani-Hamed, H.~Frost, G.~Salvatori, P.-G. Plamondon, and H.~Thomas, {\it
  {All Loop Scattering As A Counting Problem}},
  \href{http://arxiv.org/abs/2309.15913}{{\tt arXiv:2309.15913}}.

\bibitem{Arkani-Hamed:2023mvg}
N.~Arkani-Hamed, H.~Frost, G.~Salvatori, P.-G. Plamondon, and H.~Thomas, {\it
  {All Loop Scattering For All Multiplicity}},
  \href{http://arxiv.org/abs/2311.09284}{{\tt arXiv:2311.09284}}.

\bibitem{Arkani-Hamed:2023jry}
N.~Arkani-Hamed, Q.~Cao, J.~Dong, C.~Figueiredo, and S.~He, {\it
  {Scalar-Scaffolded Gluons and the Combinatorial Origins of Yang-Mills
  Theory}},  \href{http://arxiv.org/abs/2401.00041}{{\tt arXiv:2401.00041}}.

\bibitem{Arkani-Hamed:2024nhp}
N.~Arkani-Hamed, Q.~Cao, J.~Dong, C.~Figueiredo, and S.~He, {\it {Nonlinear
  Sigma model amplitudes to all loop orders are contained in the
  Tr(\ensuremath{\Phi}3) theory}},  {\em Phys. Rev. D} {\bf 110} (2024), no.~6
  065018, [\href{http://arxiv.org/abs/2401.05483}{{\tt arXiv:2401.05483}}].

\bibitem{Arkani-Hamed:2023swr}
N.~Arkani-Hamed, Q.~Cao, J.~Dong, C.~Figueiredo, and S.~He, {\it {Hidden zeros
  for particle/string amplitudes and the unity of colored scalars, pions and
  gluons}},  {\em JHEP} {\bf 10} (2024) 231,
  [\href{http://arxiv.org/abs/2312.16282}{{\tt arXiv:2312.16282}}].

\bibitem{Rodina:2024yfc}
L.~Rodina, {\it {Hidden zeros are equivalent to enhanced ultraviolet scaling
  and lead to unique amplitudes in Tr($\phi^3$) theory}},  {\em Phys. Rev.
  Lett.} {\bf 134} (2025) 031601, [\href{http://arxiv.org/abs/2406.04234}{{\tt
  arXiv:2406.04234}}].

\bibitem{Bartsch:2024amu}
C.~Bartsch, T.~V. Brown, K.~Kampf, U.~Oktem, S.~Paranjape, and J.~Trnka, {\it
  {Hidden Amplitude Zeros From Double Copy}},
  \href{http://arxiv.org/abs/2403.10594}{{\tt arXiv:2403.10594}}.

\bibitem{Li:2024qfp}
Y.~Li, D.~Roest, and T.~ter Veldhuis, {\it {Hidden Zeros in Scaffolded General
  Relativity and Exceptional Field Theories}},
  \href{http://arxiv.org/abs/2403.12939}{{\tt arXiv:2403.12939}}.

\bibitem{Zhang:2024iun}
Y.~Zhang, {\it {New Factorizations of Yang-Mills Amplitudes}},
  \href{http://arxiv.org/abs/2406.08969}{{\tt arXiv:2406.08969}}.

\bibitem{Zhou:2024ddy}
K.~Zhou, {\it {Understanding zeros and splittings of ordered tree amplitudes
  via Feynman diagrams}},  \href{http://arxiv.org/abs/2411.07944}{{\tt
  arXiv:2411.07944}}.

\bibitem{Zhang:2024efe}
Y.~Zhang, {\it {On the New Factorizations of Yang-Mills Amplitudes}},
  \href{http://arxiv.org/abs/2412.15198}{{\tt arXiv:2412.15198}}.

\bibitem{Cao:2024gln}
Q.~Cao, J.~Dong, S.~He, and C.~Shi, {\it {A universal splitting of tree-level
  string and particle scattering amplitudes}},  {\em Phys. Lett. B} {\bf 856}
  (2024) 138934, [\href{http://arxiv.org/abs/2403.08855}{{\tt
  arXiv:2403.08855}}].

\bibitem{Arkani-Hamed:2024fyd}
N.~Arkani-Hamed and C.~Figueiredo, {\it {All-order splits and multi-soft limits
  for particle and string amplitudes}},
  \href{http://arxiv.org/abs/2405.09608}{{\tt arXiv:2405.09608}}.

\bibitem{Cao:2024qpp}
Q.~Cao, J.~Dong, S.~He, C.~Shi, and F.~Zhu, {\it {On universal splittings of
  tree-level particle and string scattering amplitudes}},  {\em JHEP} {\bf 09}
  (2024) 049, [\href{http://arxiv.org/abs/2406.03838}{{\tt arXiv:2406.03838}}].

\bibitem{Stieberger:2016lng}
S.~Stieberger and T.~R. Taylor, {\it {New relations for
  Einstein\textendash{}Yang\textendash{}Mills amplitudes}},  {\em Nucl. Phys.
  B} {\bf 913} (2016) 151--162, [\href{http://arxiv.org/abs/1606.09616}{{\tt
  arXiv:1606.09616}}].

\bibitem{Schlotterer:2016cxa}
O.~Schlotterer, {\it {Amplitude relations in heterotic string theory and
  Einstein-Yang-Mills}},  {\em JHEP} {\bf 11} (2016) 074,
  [\href{http://arxiv.org/abs/1608.00130}{{\tt arXiv:1608.00130}}].

\bibitem{Chiodaroli:2017ngp}
M.~Chiodaroli, M.~Gunaydin, H.~Johansson, and R.~Roiban, {\it {Explicit
  Formulae for Yang-Mills-Einstein Amplitudes from the Double Copy}},  {\em
  JHEP} {\bf 07} (2017) 002, [\href{http://arxiv.org/abs/1703.00421}{{\tt
  arXiv:1703.00421}}].

\bibitem{Nandan:2016pya}
D.~Nandan, J.~Plefka, O.~Schlotterer, and C.~Wen, {\it {Einstein-Yang-Mills
  from pure Yang-Mills amplitudes}},  {\em JHEP} {\bf 10} (2016) 070,
  [\href{http://arxiv.org/abs/1607.05701}{{\tt arXiv:1607.05701}}].

\bibitem{delaCruz:2016gnm}
L.~de~la Cruz, A.~Kniss, and S.~Weinzierl, {\it {Relations for
  Einstein\textendash{}Yang\textendash{}Mills amplitudes from the CHY
  representation}},  {\em Phys. Lett. B} {\bf 767} (2017) 86--90,
  [\href{http://arxiv.org/abs/1607.06036}{{\tt arXiv:1607.06036}}].

\bibitem{Fu:2017uzt}
C.-H. Fu, Y.-J. Du, R.~Huang, and B.~Feng, {\it {Expansion of
  Einstein-Yang-Mills Amplitude}},  {\em JHEP} {\bf 09} (2017) 021,
  [\href{http://arxiv.org/abs/1702.08158}{{\tt arXiv:1702.08158}}].

\bibitem{Teng:2017tbo}
F.~Teng and B.~Feng, {\it {Expanding Einstein-Yang-Mills by Yang-Mills in CHY
  frame}},  {\em JHEP} {\bf 05} (2017) 075,
  [\href{http://arxiv.org/abs/1703.01269}{{\tt arXiv:1703.01269}}].

\bibitem{Du:2017kpo}
Y.-J. Du and F.~Teng, {\it {BCJ numerators from reduced Pfaffian}},  {\em JHEP}
  {\bf 04} (2017) 033, [\href{http://arxiv.org/abs/1703.05717}{{\tt
  arXiv:1703.05717}}].

\bibitem{Du:2017gnh}
Y.-J. Du, B.~Feng, and F.~Teng, {\it {Expansion of All Multitrace Tree Level
  EYM Amplitudes}},  {\em JHEP} {\bf 12} (2017) 038,
  [\href{http://arxiv.org/abs/1708.04514}{{\tt arXiv:1708.04514}}].

\bibitem{Feng:2019tvb}
B.~Feng, X.~Li, and K.~Zhou, {\it {Expansion of Einstein-Yang-Mills theory by
  differential operators}},  {\em Phys. Rev. D} {\bf 100} (2019), no.~12
  125012, [\href{http://arxiv.org/abs/1904.05997}{{\tt arXiv:1904.05997}}].

\bibitem{Zhou:2019gtk}
K.~Zhou and S.-Q. Hu, {\it {Expansions of tree amplitudes for
  Einstein\textendash{}Maxwell and other theories}},  {\em PTEP} {\bf 2020}
  (2020), no.~7 073B10, [\href{http://arxiv.org/abs/1907.07857}{{\tt
  arXiv:1907.07857}}].

\bibitem{Zhou:2019mbe}
K.~Zhou, {\it {Unified web for expansions of amplitudes}},  {\em JHEP} {\bf 10}
  (2019) 195, [\href{http://arxiv.org/abs/1908.10272}{{\tt arXiv:1908.10272}}].

\bibitem{Wei:2023yfy}
F.-S. Wei and K.~Zhou, {\it {Expanding single-trace YMS amplitudes with
  gauge-invariant coefficients}},  {\em Eur. Phys. J. C} {\bf 84} (2024), no.~1
  29, [\href{http://arxiv.org/abs/2306.14774}{{\tt arXiv:2306.14774}}].

\bibitem{Hu:2023lso}
C.~Hu and K.~Zhou, {\it {Recursive construction for expansions of tree
  Yang\textendash{}Mills amplitudes from soft theorem}},  {\em Eur. Phys. J. C}
  {\bf 84} (2024), no.~3 221, [\href{http://arxiv.org/abs/2311.03112}{{\tt
  arXiv:2311.03112}}].

\bibitem{Du:2024dwm}
Y.-J. Du and K.~Zhou, {\it {Multi-trace YMS amplitudes from soft behavior}},
  {\em JHEP} {\bf 03} (2024) 081, [\href{http://arxiv.org/abs/2401.03879}{{\tt
  arXiv:2401.03879}}].

\bibitem{Zhou:2024qwm}
K.~Zhou and C.~Hu, {\it {Towards tree Yang-Mills and Yang-Mills-scalar
  amplitudes with higher-derivative interactions}},  {\em JHEP} {\bf 01} (2025)
  167, [\href{http://arxiv.org/abs/2406.03034}{{\tt arXiv:2406.03034}}].

\bibitem{Zhou:2024qjh}
K.~Zhou, {\it {Constructing tree amplitudes of scalar EFT from double soft
  theorem}},  {\em JHEP} {\bf 12} (2024) 079,
  [\href{http://arxiv.org/abs/2406.03784}{{\tt arXiv:2406.03784}}].

\bibitem{Nicolis:2008in}
A.~Nicolis, R.~Rattazzi, and E.~Trincherini, {\it {The Galileon as a local
  modification of gravity}},  {\em Phys. Rev. D} {\bf 79} (2009) 064036,
  [\href{http://arxiv.org/abs/0811.2197}{{\tt arXiv:0811.2197}}].

\bibitem{Tseytlin:1999dj}
A.~A. Tseytlin, {\it {Born-Infeld action, supersymmetry and string theory}},
  \href{http://arxiv.org/abs/hep-th/9908105}{{\tt hep-th/9908105}}.

\bibitem{Kleiss:1988ne}
R.~Kleiss and H.~Kuijf, {\it {Multi - Gluon Cross-sections and Five Jet
  Production at Hadron Colliders}},  {\em Nucl. Phys. B} {\bf 312} (1989)
  616--644.

\bibitem{Berends:1987me}
F.~A. Berends and W.~T. Giele, {\it {Recursive Calculations for Processes with
  n Gluons}},  {\em Nucl. Phys. B} {\bf 306} (1988) 759--808.

\end{thebibliography}\endgroup

\end{document}